\documentclass[%
reprint,
superscriptaddress,
showpacs,preprintnumbers,
nofootinbib,
 amsmath,amssymb, 
 aps,
 prd,
 longbibliography,
]{revtex4-1}

\usepackage[utf8]{inputenc}

\usepackage{CJKutf8}

\usepackage{lipsum, babel}
\usepackage{cancel}
\usepackage{accents}
\usepackage{mciteplus,slashed}
\usepackage{amssymb,cancel,amsmath}
\usepackage{dcolumn}% Align table columns on decimal point
\usepackage{bm}% bold math
\usepackage[caption=false]{subfig}
\usepackage{appendix}
\usepackage{feynmp-auto}
\usepackage[T1]{fontenc}	
\usepackage{csvsimple}
\usepackage[colorlinks=true,citecolor=blue,linkcolor=blue, allcolors=blue]{hyperref}
\usepackage[capitalise]{cleveref}
\usepackage{booktabs}
\usepackage{graphicx}
\usepackage{mathrsfs}
\usepackage{appendix}
\usepackage{lmodern}

\usepackage[dvipsnames]{xcolor}
\usepackage[normalem]{ulem}

\newcommand\red[1]{\color{red}#1}
\newcommand\blue[1]{\color{blue}#1}
\newcommand\purple[1]{\color{purple}#1}
\newcommand\pinegreen[1]{\color{PineGreen}#1}
\newcommand\green[1]{\color{Green}#1}
\newcommand\cyan[1]{\color{cyan}#1}
\newcommand\magenta[1]{\color{magenta}#1}

\begin{document}

%\interfootnotelinepenalty=10000
%\interfootnotelinepenalty=\@

% PREPRINT NUMBERS
\preprint{\hfill FERMILAB-PUB-19-393-A-PPD}

\title{The High-Energy Frontier of the Intensity Frontier:\\
Closing the Dark Photon, Inelastic Dark Matter, and Muon $g-2$ Windows} %with \\ Existing Probes and the LongQuest Proposal}

\author{Yu-Dai Tsai}
\email{ytsai@fnal.gov}
\affiliation{Fermilab, Fermi National Accelerator Laboratory, Batavia, IL 60510, USA}
\affiliation{Kavli Institute for Cosmological Physics, University of Chicago, Chicago, IL USA}
\author{Patrick deNiverville}
\email{pgdeniverville@gmail.com}
\affiliation{Center for Theoretical Physics of the Universe, IBS, Daejeon 34126, Korea}
\author{Ming Xiong Liu}
\email{ming@bnl.gov}
\affiliation{Los Alamos National Laboratory, Los Alamos, NM 87545, USA}

\date{\today}

\begin{abstract}

We study hidden sector and long-lived particles at past (CHARM and NuCal), present (NA62 and SeaQuest/DarkQuest), and future (LongQuest) experiments that are at the high-energy frontier of the intensity frontier. We focus on exploring the minimal vector portal and variere-lifetime particles (VLP). VLP models have mostly been devised to explain experimental anomalies while avoiding existing constraints, and we demonstrate that proton fixed-target experiments provide one of the most powerful probes for the sub-GeV to few GeV mass range of the VLP models, using inelastic dark matter (iDM) as an example. We consider an iDM model with small mass splitting that yields the observed dark matter (DM) relic abundance, and a scenario with a sizable mass splitting that can also explain the muon $g-2$ anomaly. We set strong limits based on the CHARM and NuCal experiments, which come close to excluding iDM as full-abundance thermal DM candidates in the MeV to GeV mass range, for the mass arrangements and small mass splittings we consider. We also study the future projections based on NA62 and SeaQuest/DarkQuest, and update the constraints of the minimal dark photon parameter space. We found that NuCal sets the only existing constraint in $\epsilon \sim 10^{-8} - 10^{-4}$ regime reaching $\sim$ 800 MeV in dark photon mass due to the resonant enhancement of the proton bremsstrahlung production. Finally, we propose LongQuest, a three-stage thorough retool of the SeaQuest experiment with short ($\lesssim$ 5 m), medium ($\sim$ 5 m), and long baseline ($\gtrsim$ 35 m) tracking stations/detectors, as a multi-purpose machine to explore dark sector particles with a wide range of couplings to the standard model sector.

\end{abstract}

\maketitle

\section{Introduction}

The search for particles that can account for the dark matter (DM) abundance or mediate the interactions between DM and Standard Model (SM) particles is one of the central topics of modern particle physics \cite{Cushman:2013zza,Essig:2013lka}. 
The most interesting search targets are particles capable of explaining both the DM relic abundance and other experimental or observational anomalies. 
In recent years, given the non-observation of new physics at the TeV scale thus far and the ever-stronger constraints on Weakly Interacting Massive Particle (WIMP) DM \cite{Cushman:2013zza}, many particle physicists have shifted their attention to the study of dark sector particles with sub-GeV to few GeV masses \cite{Alexander:2016aln,Battaglieri:2017aum}. The MeV-GeV energy regime is of great interest due to a number of experimental anomalies, including the muon $g-2$ anomaly \cite{Bennett:2006fi}, the proton charge radius anomaly (see, e.g. \cite{Pohl:2010zza}), and the LSND/MiniBooNE excess (see, e.g. \cite{Athanassopoulos:1995iw,Aguilar-Arevalo:2018gpe}). 
New physics models with small couplings to the SM sector have been proposed to explain these anomalies.
These dark sector particles could hide behind the SM backgrounds and can only be observed in high-intensity environments.
The small couplings also result in long lifetimes and can lead to displaced vertices and other unconventional signatures at the LHC and other colliders. These searches are thus called long-lived particle (LLP) searches (see, e.g. \cite{Alimena:2019zri}).

Proton fixed-target machines are among the most powerful and robust probes of dark sector and LLPs in the MeV to $\sim$10 GeV regime (see, e.g., \cite{deNiverville:2011it, Kahn:2014sra,Gardner:2015wea,deNiverville:2016rqh, Izaguirre:2017bqb,Pospelov:2017kep,Magill:2018jla,Berlin:2018pwi,Magill:2018tbb,Arguelles:2018mtc,Arguelles:2019xgp}). Firstly, one benefits from the combination of high energy and high intensity that the proton beams provide. Secondly, these accelerator-based probes (same for the LHC) do not depend on DM abundance, velocity distribution (comparing to DM direct detection, e.g., \cite{Cushman:2013zza}), or cosmic history  (comparing to cosmological constraints, e.g., \cite{Boehm:2013jpa,Alexander:2016aln}). The constraints and sensitivity reach also do not depend on complicated astrophysics or rare events (see, e.g., supernova constraints \cite{Raffelt:1996wa,Chang:2016ntp,Hardy:2016kme,Bar:2019ifz}).
As our theoretical understanding of the dark sector advances, it is crucial to look back at the highest-energy experiments of the intensity frontier. These include past experiments like CHARM, NuCal ($\nu$-Cal I), and ongoing experiments like NA62, SeaQuest/SpinQuest, and their beam-dump/upgrade proposals (the NA62 beam-dump run and the installation of a new ECal at SeaQuest/SpinQuest).
It is also interesting to look more closely at these existing facilities to see if further upgrades can be undertaken to improve the sensitivity, as will be discussed in this paper.

The most common LLP searches focus on minimal models with couplings to the SM through renormalizable interactions ("portals"), which includes the vector, Higgs, and neutrino portals (see, e.g. \cite{Alexander:2016aln,Battaglieri:2017aum} for detailed descriptions). We are interested in the vector portal in the MeV to GeV regime, in which many experiments are considered and proposed in search of the signatures associated with this portal \cite{Okun:1982xi,Galison:1983pa,Holdom:1985ag,Pospelov:2007mp,ArkaniHamed:2008qn,Bjorken:2009mm,Aubert:2009cp,Curtin:2013fra,Lees:2014xha,Ablikim:2017aab,Aaij:2017rft,Archilli:2011zc,KLOE:2016lwm,Bergsma:1985is,Bergsma:1985qz,Bernardi:1985ny,Konaka:1986cb,Riordan:1987aw,Bjorken:1988as,Bross:1989mp,Davier:1989wz,Blumlein:1990ay,Blumlein:1991xh,Athanassopoulos:1997er,Astier:2001ck,Bjorken:2009mm,Essig:2010gu,Williams:2011qb,Gninenko:2011uv,Abrahamyan:2011gv,Blumlein:2011mv,Merkel:2011ze,Gninenko:2012eq,Blumlein:2013cua,Andreas:2012mt,Merkel:2014avp,Bernardi:1985ny,MeijerDrees:1992kd,Archilli:2011zc,Gninenko:2011uv,Babusci:2012cr,Adlarson:2013eza,Agakishiev:2013fwl,Adare:2014mgk,Batley:2015lha,KLOE:2016lwm,Essig:2010xa,Freytsis:2009bh,Balewski:2013oza,Wojtsekhowski:2012zq,Beranek:2013yqa,Raggi:2014zpa,Echenard:2014lma,Battaglieri:2014hga,Alekhin:2015byh,Gardner:2015wea,Ilten:2015hya,Curtin:2014cca,He:2017ord,Kozaczuk:2017per,Ilten:2016tkc,Feng:2017uoz,Alexander:2017rfd,Pospelov:2017kep}. We revisit several of the existing constraints, fill in the missing relevant production channels, and conduct a robust reanalysis based on the available data.

In this paper, we also focus on a search for a next-to-minimal class of models, called variere-lifetime particles (VLP) search. We loosely define VLP as next-to-minimal dark-sector model that the production and decay (signature) involve different physics or distinctive parameters and can thus be treated as independent\footnote{Means one can fix a process while vary the other. VLP has a definition that is accelerator-probe centric, just like LLP is an LHC/collider-centric description and many of the LLPs are actually too {\it short lived} for many fixed-target experiments given the long baselines.}.
In many situations, VLP are considered to have a certain range of lifetime or decay length to avoid experimental or observational bounds while explaining specific anomalies. Thus they can be either long-lived or short-lived with respect to the experimental baseline or detector size.
Examples of VLP include inelastic dark matter (iDM) (see, e.g. \cite{TuckerSmith:2001hy,Bramante:2016rdh,Izaguirre:2017bqb,Giudice:2017zke,Berlin:2018jbm,Mohlabeng:2019vrz,Eby:2019mgs}) and dark neutrinos\footnote{Heavy sterile neutrinos that not only have mixing with SM neutrinos, 
but also charged under dark $U(1)_D$ in an anomaly-free fashion. Proposed to explain the MiniBooNE anomaly \cite{Bertuzzo:2018itn,Ballett:2018ynz,Arguelles:2018mtc}}
\cite{Bertuzzo:2018itn,Ballett:2018ynz,Arguelles:2018mtc}. In the case of iDM, the production of the dark sector particles is predominantly through small SM coupling, but the decay length of the dark sector particles in the search is determined by the mass-splitting of nearly degenerate dark particle states (see Sec. \ref{sec:model} for more detailed discussions).

iDM, from its formulation, was developed to explain the DAMA anomaly while avoiding the constraints from other experiments \cite{Abusaidi:2000wg,Bernabei:2000qi,TuckerSmith:2001hy}. iDM also avoids strong constraints like those from the Cosmic Microwave Background (CMB) \cite{Ade:2015xua,Slatyer:2015jla} by heavily suppressing the dark matter (co-)annihilation cross section. iDM thus provides one of the few viable GeV/sub-GeV thermal DM scenarios that freezes out to the right relic abundance, also called "thermal targets", since many future experiments are proposed to explore these models (see, e.g. \cite{Izaguirre:2015yja,Alexander:2016aln}). The thermal targets include freeze-in DM \cite{Moroi:1993mb,Asaka:2005cn,Shaposhnikov:2006xi,Kusenko:2006rh,Hall:2009bx}; asymmetric DM \cite{Kaplan:2009ag}; DM annihilating into scalar or pseudoscalar mediators \cite{Krnjaic:2015mbs,Dolan:2014ska};  SIMP/ELDER DM \cite{Hochberg:2014dra,Hochberg:2015vrg,Kuflik:2015isi,Kuflik:2017iqs}. Descriptions of these models and a compilation of the probes can be found in \cite{Battaglieri:2017aum}.
Recently, iDM was also studied as the last standing viable dark photon explanation of the muon $g-2$ anomaly \cite{Bennett:2006fi,Pospelov:2008zw,Mohlabeng:2019vrz}. iDM with a sizable mass splitting $\Delta$ was considered in \cite{Mohlabeng:2019vrz}, causing semi-visible decays that avoid both visibly-decaying dark photon searches and the usual invisible dark photon constraints \cite{Banerjee:2016tad,Lees:2017lec}.

In this paper, we will demonstrate that proton fixed target and beam dump experiments provide some of the strongest constraints on iDM in the MeV to GeV regime.
Combined with a recent study on dark neutrinos \cite{Arguelles:2018mtc}, we demonstrate a general point that high-energy proton fixed target machines provide strong probes of VLP models that would otherwise escape the bounds from other experiments.
Throughout our analysis, we conduct simplified detector simulations which provide a reasonable estimate of the experimental event rate.

Firstly, we set the strongest constraints on iDM model parameters in the MeV to GeV regime, based on CHARM and NuCal beam-dump experiments, for both small and sizable iDM mass splitting. We rule out a large portion of the parameter space for iDM to account for the total DM abundance.
Beyond the new iDM bounds, we also revisit the minimal dark photon bounds set by the pioneering analyses \cite{Blumlein:2011mv,Gninenko:2012eq,Blumlein:2013cua}. We include several additional but relevant production channels and conduct a simplified detector simulation (which is also an improvement from the previous treatments) to achieve robust re-evaluation of the constraints.

Secondly, we perform the first iDM sensitivity projection based on the NA62 beam-dump run, as well as a dark photon analysis\footnote{Also, see \cite{Dobrich:2018ezn} for NA62's conference proceedings.}. NA62 can further extend the sensitivity of the two scenarios and help to close the window of parameter space for which iDM models can explain the muon $g-2$ anomaly.

Furthermore, we conduct a study of iDM in the muon $g-2$ regime based on the DarkQuest upgrade of the SeaQuest experiment \cite{SeaQuest_slides,Gardner:2015wea,Berlin:2018pwi}. We find that the current DarkQuest configuration does not improve the sensitivity in the iDM muon $g-2$ regime. 
Thus, we consider the possibilities of extending the available decay volume, turning off or tuning the KMAG magnetic field, and adding additional detectors stations to extend the sensitivity of both iDM and dark photon searches. 

Inspired by the studies of SeaQuest/DarkQuest, we consider a complete retool of the SeaQuest/SpinQuest experiment as a dedicated and multi-purpose experiment to conduct dark sector searches with short ($\lesssim$ 5 m), medium ($\sim$ 5 m -- 12 m), and long-baseline ($\gtrsim$ 35 m) detectors. One could improve the primary decay detector (installing a gas or hadron-blind detector), install another new ECal detector in the open space behind station 4, host an additional exotic millicharged-particle detector \cite{Kelly:2018brz}, and adding a new beam-dump plus a tracking detector in front of the FMAG (to help the prompt dimuon study). 
For convenience, we group all these new installations as "LongQuest" and will discuss this three-stage retool in Sec. \ref{sec:LongQuest}.

The paper is organized as follows: in Sec. \ref{sec:model}, we introduce dark photon and iDM models. In Sec. \ref{sec:experiments}, we describe our analysis and detail the experimental configurations of CHARM, NuCal, NA62, SeaQuest/DarkQuest, and LongQuest. In Sec. \ref{sec:discussion} we discuss the results and implications, and we conclude by discussing possible future directions of inquiry in Sec. \ref{sec:future}. 
Before getting into details, we provide a summary of information on the experiments studied in Table \ref{table:1} below.
\begin{table}[h!]
\centering
\begin{tabular}{|c | c | c | c| c|} 
 \hline
 Experiment & Beam Energy & POT & $L_{\rm dist.}$ & $L_{\rm dec}$ \\ [0.5ex] 
 \hline
CHARM & 400 GeV & 2.4e18 & 480 m & 35 m \\  \hline
NuCal & 70 GeV   & 1.7e18 & 64 m & 23 m \\ \hline
NA62 & 400 GeV & *1.3e16/1e18 & 82 m & 75 m  \\ \hline
SQ/DQ & 120 GeV & *1.4e18/1e20 & 5 m & *7 m \\ \hline
LongQuest & 120 GeV & *1e20 & 5 m & *7/13 m \\ [1ex] 
 \hline
\end{tabular}
\caption{This table provides a comparison of experiments considered in this paper. *Indicates not yet decided; $L_{\rm dist.}$ is the distance from the target to the decay region; $L_{\rm dec.}$ is the fiducial particle decay length. The detector areas $A_{\rm dec.}$ are more complicated and not listed in the table. Our information regarding the NA62 experimental configuration was updated directly through contact with the NA62 collaboration \cite{PrivateCom}.}
\label{table:1}
\end{table}

\section{Models}\label{sec:model}

In this section, we will outline two specific models studied in this paper. We will focus on the visibly decaying dark photon produced through kinetic mixing with the photon, and inelastic dark matter (iDM) models consisting of a nearly mass-degenerate pair of dark sector fermions. We will also briefly discuss an iDM model with a sizable mass splitting that can explain the muon $g-2$ anomaly while avoiding existing experimental constraints.

\subsection{Dark Photon from Kinetic Mixing}

We first introduce a minimal model of a visibly decaying dark photon through the kinetic mixing between SM U(1) hypercharge and a dark U(1) gauge field. The Lagrangian can be written as: 
\begin{equation}
\mathcal{L}_\mathrm{kin.\;mix.} = \frac{\epsilon}{2\cos\theta_W}
A^{'}_{\mu\nu} B^{\mu\nu},
\end{equation}
where $\theta_W$ is the Weinberg angle. The crucial interaction term between dark photon and SM particles can be expressed as: 
\begin{equation}
\mathcal{L}_\mathrm{\rm int} \supset 
\epsilon e A^{'}_\mu \mathcal{J^\mu_{\rm EM}},
\end{equation}
where $\mathcal{J^\mu_{\rm EM}}$ is the SM electromagnetic current.

We consider the case of a massive $A'$ where the mass $m_{A^{'}}$ can be generated through the Higgs or Stueckelberg mechanisms. Our bounds and sensitivity projections are insensitive to the mass-generation mechanism.

There are three main production channels considered in the literature for kinetically-mixed dark photons. 
\begin{itemize}
    \item Meson decays
    \item Proton bremsstrahlung
    \item Drell-Yan and QCD processes
\end{itemize}
Each of these processes dominates dark photon production for different dark photon masses.
Among these, the Drell-Yan and QCD production processes suffer large uncertainties in the sub-GeV energy regime given the large uncertainty in the parton distribution functions (see, e.g., \cite{Ball:2012cx,Feng:2017uoz}). Thus, although these processes are in principle important for dark sectors, especially for above GeV-scale masses, we do not include them to provide a conservative production estimate.

All the other relevant production processes were considered, as discussed in Sec. \ref{sec:experiments}. 
In Sec. \ref{sec:discussion_dp}, we present updates of fixed-target constraints from CHARM and NuCal, taking into account some relevant production channels neglected in the literature. We also present the sensitivity projections of NA62 beam-dump mode and SeaQuest/DarkQuest upgrades.

\subsection{Inelastic Dark Matter} \label{sec:iDM_g-2}

We focus on one of the simplest incarnations of iDM in this article. We study iDM composed of a Dirac pair of two-component Weyl spinors, $\eta$ and $\xi$, charged under a new $U(1)$ gauge symmetry $U(1)_D$. For the convenience of comparing to previous results, we mostly follow the notation of Ref. \cite{Berlin:2018pwi}. Other similar models that are also referred to as iDM (see \cite{TuckerSmith:2001hy} and the follow-ups) can also be probed by the experimental techniques considered here.
We include the dark matter interaction term as
\begin{equation}
\mathcal{L}_\mathrm{\rm int} \supset  \epsilon e A^{'}_\mu \mathcal{J^\mu_{\rm EM}}+ g_D A^{'}_\mu 
 \mathcal{J^\mu_{\rm D}}.
\end{equation}
$\mathcal{J^\mu_{\rm D}}$ is the dark sector current to which the dark photon $A^{'}$ couples. Similarly to the definition of the electromagnetic fine structure constant, we express 
$g_{D} \equiv \sqrt{4\pi \alpha_{D}}$.
For the model considered, the dark sector current consists of a four-component fermionic state $\Psi=(\eta\;\xi^{\dagger})$ with two-component Weyl spinors $\eta$ and $\xi$.
Here again, $A'$ is massive and $U(1)_D$ is broken. One can write down the Majorana mass terms as well as the Dirac mass term as:

\begin{equation}
\mathcal{L} \supset - m_D \eta \xi - \frac{1}{2} \delta_\eta \eta^2 - \frac{1}{2} \delta_\xi \xi^2 +\rm{h.c.}
\end{equation}
$\delta_\eta,\delta_\xi \ll m_D$ are technically natural values for the parameters because they break the $U(1)$ symmetry explicitly. 
After the mass diagonalization, $m_{12} \simeq m_D \mp \frac{1}{2}(\delta_{\eta}+\delta_{\xi}) $, we have 
$\chi_1\simeq i(\eta-\xi)/\sqrt{2}, \chi_2\simeq (\eta+\xi)/\sqrt{2}.$

We can express the relevant parts of the Lagrangian in terms of the mass eiganstates $\chi_1$ and $\chi_2$ as
\begin{equation}
\mathcal{L} \supset \sum_{i = 1, 2} \bar \chi_i (i \slashed{\partial} - m_{\chi_{i}}) \chi_i - ( {g_D} A'_\mu \bar{\chi_1} \gamma^{\mu}\chi_2 + \rm{h.c.}).
\end{equation}
The elastic interactions are suppressed by a factor of $\delta/m_D.$ $\delta\ll m_D$ is again technically natural because the $U(1)$ explicit breaking would be restored when $\delta\rightarrow 0.$ Note that the elastic interaction vanishes as $\delta_\eta=\delta_\xi.$

The particle $\chi_1$, which we take to be lighter than $\chi_2$, can account for the current-day dark matter abundance. The mass splitting is defined as:
\begin{equation}
\Delta \equiv \frac{m_2-m_1}{m_1}.
\end{equation}

We take $\Delta\ll 1$ as we can assume $\delta_{\eta,\xi}\ll m_D$ with the same technical naturalness argument. One of the main goals of this paper is to exclude the canonical iDM with small mass-splitting $\Delta,$ but in the next section, we will also consider a regime of iDM parameter space in which $\Delta \gtrsim 0.4$ while elastic coupling remains small, tuned to explain the muon $g-2$ anomaly while avoiding current experimental bounds. 
Even though the consideration of large $\Delta$ values seems theoretically arbitrary, this regime of parameter space is of great interest, because it can simultaneously explain the muon $g-2$ anomaly and provide the correct DM relic abundance \cite{Fayet:2004bw,Izaguirre:2015yja}. The calculation of the dark matter relic abundance is detailed in, e.g., Ref. \cite{Izaguirre:2017bqb}.

One can write down an approximate analytical expression for the $A^{'}$ decay width when the mass splitting $\Delta$ and the elastic coupling terms are small, as $A^{'}$ dominantly decays to $\chi_1 \chi_2$ with a width:
\begin{equation}
\Gamma(A^{'}\rightarrow \chi_1 \chi_2)\simeq \frac{\alpha_D m_{A^{'}}}{3}
\sqrt{1-\frac{4 m_1^2}{m^2_{A^{'}}}}\left(1+\frac{2m_1^2}{m^2_{A^{'}}}\right)
\end{equation}

An approximate expression for the width of the $\chi_2$ decay exists in the small $\Delta$ limit when $m_{A^\prime}\gg m_1 \gg m_l.$
\begin{equation}
\Gamma(\chi_2\rightarrow\chi_1 l^{+} l^{-}) \simeq \frac{4\epsilon^2\alpha_{\rm em}\alpha_D \Delta^5 m_1^5}{15\pi m_{A^{'}}^4}
\end{equation}

In our analysis, we simulated the three-body decay $\chi_2 \to e^+ e^- \chi_1$ by sampling the full decay width, calculated in the Appendix \ref{app:chi2dec}, rather than using the approximation. The dilepton signal provides the most viable signature of iDM in a beam-dump type decay experiment. 
Further details on the handling of the production and detection can be found in Sec. \ref{sec:experiments}.

\subsection{Closing the iDM Thermal Target and Muon $g-2$ Window}

The muon magnetic moment can be written as $a_\mu \equiv (g_\mu-2)/2,$ with $g_\mu$ being the the muon g-factor \cite{Bennett:2006fi,Tanabashi:2018oca}.

The difference between the experimentally measured value of $a_\mu$ and that predicted in the SM is:
\begin{equation}
\Delta a_\mu = a^{\rm exp}_\mu - a^{\rm th}_\mu = 268 (63)(43) \times 10^{-11}, \end{equation} 
indicating a 3.5 $\sigma$ discrepancy between the theoretical prediction and the experimental measurement \cite{Tanabashi:2018oca}. The ongoing Fermilab muon $g-2$ experiment \cite{Grange:2015fou} and J-PARC muon $g-2$/EDM experiment \cite{Abe:2019thb} will reduce the current experimental uncertainty. Progress has been made in reducing the theoretical uncertainty in SM prediction in Refs. \cite{Blum:2018mom,Izubuchi:2018tdd,Keshavarzi:2018mgv}.

A dark photon has been proposed as a possible explanation of the muon $g-2$ anomaly \cite{Fayet:2004bw,Pospelov:2008zw}.
The minimal model assumes that the dark photon either decays visibly \cite{Bross:1989mp,Bjorken:1988as,Riordan:1987aw,Lees:2014xha,Batley:2015lha,Merkel:2014avp} or has invisible decay channels to dark matter particles that could account for the current-day relic abundance \cite{Fayet:2004bw,Izaguirre:2015yja}, but both of these possibilities have been excluded by various experiments \cite{Batley:2015lha,Banerjee:2016tad,Lees:2017lec}. %Mami/NA64/BaBar
These constraints can be weakened if the dark photon is allowed to decay semi-visibly, as is possible in iDM models \cite{Mohlabeng:2019vrz}. If $\Delta$ is sufficiently large, the  $\chi_2$ will decay inside the detector and thus avoid the invisible decay bounds on elastic dark matter (e.g., the \textsc{BaBar} monophoton bound \cite{Lees:2017lec} and NA64 \cite{Banerjee:2016tad}).

In this paper, we present a detailed discussion of the sensitivity of proton-beam fixed target facilities to iDM with $\Delta \gtrsim 0.4$, consisting a parameter regime that could explain the muon $g-2$ anomaly and avoid other experimental constraints. We will discuss the results in Sec. \ref{sec:discussion_iDM}.
We found that the existing fixed-target beam dump data can set strong limits on the $\Delta=0.4$ iDM parameter space, but cannot constrain the muon $g-2$ target regime because the distances between the targets and the decay regions are too large, and thus $\chi_2$ do not survive to reach the fiducial decay region. In the case of SeaQuest/DarkQuest (SQ/DQ), efforts to reduced SM backgrounds (such as the KMAG magnetic field in SQ/DQ) suppresses the number of charged decay products able to reach the detector.
One could consider NA62 beam dump mode or further upgrades of SQ/DQ (e.g., our LongQuest proposal) to improve the sensitivity in this regime, as will be discussed in the next section.

We present several benchmark iDM parameter sets as detailed below. We consider fermionic iDM and fix the mediator to dark matter mass ratio to be $m_{A^\prime} = 3 m_1.$
In the small mass splitting regime, we consider a fractional mass difference $\Delta$ = 0.05 with $\alpha_D$ = 0.5. Also, we fix the value of the kinetic mixing $\epsilon$ such that the model yields the correct relic abundance while producing the observed relic abundance. We also show these for the case of $\Delta$ = 0.1, $\alpha_D$ = 0.1, and then fix the $\epsilon$ to that of relic abundance and varies $\alpha_D$.
We then consider the muon $g-2$ motivated regime for iDM with a sizable mass splitting regime $\Delta$ = 0.4 and coupling $\alpha_D$ = 0.1. Then, we fix the $\epsilon$ to the value that gives the correct muon $g-2$ value while varying $\alpha_D$ to compare the current constraints and future sensitivity reach.
We show the new constraints and projections for the small mass-splitting iDM parameter space in Fig. \ref{fig:1_iDM} and the muon $g-2$ motivated regime in Fig. \ref{fig:iDM_g-2}.

\section{Analysis and Experimental Configurations}
\label{sec:experiments}

The general search strategy is to look for the decay of the dark photon through $A^\prime \to \ell^+ \ell^-$ or the three-body semi-visible decay $\chi_2\to\chi_1 \ell^{+} \ell^{-}$. The decay signature is a lepton-antilepton pair, and the best probes are "decay detectors" (discussed in Sec. \ref{sec:future}) in a fixed target setup. In this section, we discuss each of the experiments considered, including CHARM, NA62, NuCal/U70, SeaQuest/DarkQuest, and our proposed LongQuest setup.
In Table \ref{table:1}, we summarize some useful experimental information for comparison.

We simulated the production, propagation, and decay of dark sector particles using a variant of the \textsc{BdNMC} code \cite{deNiverville:2016rqh}. Production via radiative $\pi^0$ and $\eta$ decays and dark photon bremsstrahlung was considered for each of the experiments. The productions from other mesons are estimated and found to be subdominant and negligible for the probes and processes we consider.

Meson decay requires careful handling, as the calculation of the dark photon or dark matter decay spectrum from meson decays requires knowledge of the angular-momentum distribution of the $\pi^0$ and the $\eta$. As neutrino and beam dump experiments rarely consider the $\pi^0$ or $\eta$, we must either use distributions generated with Monte Carlo tools or meson production distributions (see Ref. \cite{Dobrich:2019dxc} for a helpful review of existing pseudoscalar meson production data and how closely it agrees with simulations and production distributions). $\pi^0$ and $\eta$ production distributions are not widely available, but previous experimental studies \cite{Amaldi:1979zk,Jaeger:1974pk} provide strong evidence that the average of a $\pi^+$ and $\pi^-$ distribution closely resembles the $\pi^0$ distribution and when well above the production energy threshold, the $\eta$ distribution. Unless otherwise stated, we adopt the charged pion distributions of Ref. \cite{BMPT:Bonesini:2001iz} to describe both $\pi^0$ and $\eta$ production and hereafter we will collectively refer to them as the BMPT distribution. Mesons are generated from the BMPT distribution using an acceptance-rejection algorithm, and due to their extremely short lifetime of $<10^{-17}\,\mathrm{s}$, are assumed to decay without propagating any significant distance. In the case of a visibly decaying dark photon, we consider only the on-shell process $\pi^0,\eta \to \gamma \gamma^\prime$, while for iDM we also consider the off-shell process $\pi^0,\eta \to \gamma \gamma^{\prime*} \to \gamma \chi_1 \chi_2$.

Production through dark photon bremsstrahlung was simulated by sampling the production cross-section calculated using the Improved Weizacher-Williams (IWW) approximation \cite{Blumlein:2013cua,Gorbunov:2014wqa,deNiverville:2016rqh} with an acceptance-rejection algorithm. Note that the use of the IWW approximation requires us to put limits on $z\equiv E_{A'}/E_\mathrm{Beam}$ and on the transverse momentum of the $A^\prime$, $p_T$, to remain in the regime where the approximation is valid. For all of the experiments discussed below, we limit $z\in [0.1,0.9]$ and $p_T<1\,\mathrm{GeV}$.

The result of our production simulation is a list of four-momenta for the long-lived particle $y = \gamma^\prime, \chi_2$, depending on the model studied. A given particle $y_i$ has a probability of decaying visibly in a given detector equal to

\begin{equation*}
\label{eq:dec_prob}
 P_\mathrm{decay,i} = \mathrm{Br}_{X+\ell^+\ell^-} \left[\exp\left(-\frac{L_{1,i} E_i}{c\tau m_{\alpha}}\right) - \exp\left(-\frac{L_{2,i} E_i}{c\tau m_y}\right)\right],
\end{equation*}
where
\begin{itemize}
    \item $\mathrm{Br}_{X+\ell^+\ell^-}$ is the branching ratio of $y \to X+\ell^+\ell^-$.
    \item $E_i$ and $\tau$ are the energy and lifetime of the $y_i$, respectively.
    \item $L_{1,i}$ ($L_{2,i}$) is the distance that $y$ propagates before entering (exiting) the decay volume.
\end{itemize}

Each $y$ is decayed into an $X+\ell^+\ell^-$ end state. In order for an event to be accepted, the resulting leptons must satisfy experiment-specific cuts on their energy and propagation direction. The total event rate from a given channel can then be written as
\begin{equation}
 \label{eq:decay_event_rate}
 N_{\mathrm{event},j} = \frac{N_j \epsilon_\mathrm{eff}}{N_\mathrm{trials}} \sum_i P_\mathrm{decay,i} \theta(p_{e^+,i},p_{e^-,i}),
\end{equation} where \begin{itemize}
 \item $N_j$ is the total number of $y$ produced through a given production channel. This is equal to $N_{\pi^0,\eta} \times \mathrm{Br}(\pi^0,\eta \to X+y)$ for the meson decay channels, and $N_\mathrm{brem}= \mathrm{POT} \times \sigma_\mathrm{brem}/\sigma_{pp-\mathrm{inelastic}}$ for dark photon bremsstrahlung.
 \item $P_\mathrm{decay,i}=0$ if the $y$ does not intersect the detector,
 \item $\theta(p_{e^+,i},p_{e^-,i})=1$ if the end-state leptons satisfy the experiment-specific cuts and 0 otherwise,
 \item $\epsilon_\mathrm{eff}$ is the detection efficiency, and
 \item $N_\mathrm{trials}$ is the total number of $y$ trajectories generated.
\end{itemize} The total event rate is found by summing over all production channels $j$.

\subsection{CERN-Hamburg-Amsterdam-Rome-Moscow (CHARM) Experiment}

The CHARM experiment with its decay detector (DD) has been one of the most potent high-energy proton fixed target machines to explore the dark sector. A detailed description of the CHARM experiment and the CHARM DD can be found in Ref. \cite{Bergsma:1983rt}.

The CHARM DD is located 480 meters downstream of the target and consists of a cuboid decay volume with dimensions $3 \times 3 \times 35 \rm\;m^3$ followed by a calorimeter. There are three chamber modules inside the decay volume to track charged particles. One feature of the CHARM DD is that it is offset from the neutrino beamline by a distance of 5 meters, which corresponds to an off-axis angle of 10 mrad.

As mentioned above, mesons were generated using the BMPT distribution, with the number of $\pi^0$'s produced per Proton on Target (POT) estimated to be 2.4 \cite{deNiverville:2018hrc}. The $\eta$ production rate per POT is scaled to the $\pi^0$ rate, with $N_\eta = 0.11\times N_{\pi^0}$.

In our analysis, we follow the analysis cuts discussed in Refs. \cite{Bergsma:1983rt,Gninenko:2012eq} and run our own Monte Carlo (MC) production and simplified detector simulations. The $\chi_2$ decays are required to occur in the decay volume (no hits in the scintillator plane (SC1) in front of the decay volume) and must be read out in the calorimeter (hits in at least 4 scintillator planes (SC2) of the calorimeter module) located at the end of the 35 meter decay volume, with a total shower energy greater than 3 GeV. The detection efficiency of these events is assumed to be 0.6. The data sample was scanned in search of events as described in \cite{Bergsma:1983rt}. The background events consist of tracks from neutrino interactions, beam-produced muons, and cosmic rays were filtered out from their final event sample through the cuts. No events were found satisfying the selection criteria listed in the experimental data.

The fine-grained calorimeter attached to the neutrino detector used in the wide-band neutrino beam experiment could also provide sensitivity to new physics \cite{Dorenbosch:1986nh}. We expect the sensitivity to iDM and dark photons to be worse because of its much smaller decay volume and larger background. However, the CHARM neutrino detector can potentially provide sensitivity to different signatures (e.g., scattering with SM particles), providing sensitivity to other dark sector scenarios.  

\subsection{$\nu$-Calorimeter I (NuCal) Experiment}

Next, we focus on the analysis of the $\nu$-calorimeter I ($\nu$-CAL I, simplified as NuCal) neutrino experiment.
NuCal took data from the U70 accelerator's 70 GeV proton beam impacting onto an iron target. The details of the experiments can be found in ref. \cite{Blumlein:1990ay} and previous searches for dark sector particles can be found in refs. \cite{Blumlein:1990ay,Blumlein:1991xh,Blumlein:2011mv,Blumlein:2013cua}. 
NuCal collected $N_{\rm tot} = 1.71 \times 10^{18} $ POT during the beam-dump run. We estimated $N_{\pi^0}/\mathrm{POT}=1.96$ by comparing $\sigma(pp \to \pi^0 + X)$ \cite{Blumlein:2011mv} to  the inelastic $pp$ cross section. $N_\eta$ was assumed to be approximately $0.11\times N_{\pi^0}$. We considered a cylindrical fiducial decay volume with a length of 23.0 m and a diameter of 2.6 m, beginning 64.0 meters downstream of the beam-dump target \cite{Blumlein:2011mv}. 
We focus on deriving new bounds on iDM in the small mass-splitting and muon $g-2$ inspired regimes and revisit the dark photon analysis by adding the $\eta$-meson production channel, which was absent in the previous analyses.

For our iDM analysis, we utilize the cuts from ref. \cite{Blumlein:2011mv}.
We require a minimal electromagnetic (EM) shower energy of $E_\mathrm{em}$ > 3 GeV and that each member of the end state lepton pair possess a maximal angle with respect to the beam direction of $\theta_{\rm em}$ < 0.05 rad.
An additional cut on the maximum hadronic shower energy of $E_\mathrm{had} < 1.5$ GeV was also imposed on the data but did not impact our simulation. There were 3880 reconstructed events from the $N_{\rm tot} = 1.71 \times 10^{18} $ POT run, and 5 events survived the aforementioned cuts, while the background estimation from the simulation of neutrino interactions predicted 3.5 SM events. We exclude model parameter space that predicts more than 6.06 events at a $90\%$ confidence level, with a reconstruction efficiency of $\epsilon=70\%$ \cite{Blumlein:1990ay}. 
%6.06 events for this one.
We use a 3 GeV cut for the iDM analysis instead of the lower background 10 GeV cut adopted for our dark photon analysis because the electron pairs from the semi-visible decays are too soft to survive the cut reliably.

Our dark photon analysis employed the selection cuts considered in Refs. \cite{Blumlein:1990ay,Blumlein:2013cua} with the same $N_{\rm tot}=1.7 \times 10^{18}$ POT data taken. 
For the case where dark photons decay to electrons, the event signature is a single EM shower along the direction of the beam. An energy cut of $E > 10\,\mathrm{GeV}$ is imposed for the EM showers and the efficiency is reported as $\epsilon = 70\%$ \cite{Blumlein:1990ay}. 
Among the 3880 reconstructed events, only one isolated SM shower survived the selection cuts. One observed event is consistent with the neutrino background simulation that predicts 0.3 events from SM sources.
For a 90 percent confidence level exclusion we require $n_{\rm event}$=3.64 events from dark photon decays (see \cite{Tanabashi:2018oca}).

\subsection{NA62 Experiment}

The NA62 experiment utilized the 400 GeV Super Proton Synchrotron (SPS) beam at CERN to measure rare, charged kaon decays. Information on the NA62 setup and a proposed beam-dump run can be found in Refs. \cite{NA62:2017rwk,Dobrich:2018ezn,Dobrich:2019dxc}. It is worth noting that there are other dark photon studies, utilizing rare charged kaon and charged pion leptonic decays \cite{Chiang:2016cyf} or studying $K^{+} \rightarrow \pi^+\pi^{0}$ (followed by $\pi^0 \rightarrow A^{'}\gamma$) \cite{CortinaGil:2019nuo}, that are completely different analyses and cover different regimes comparing to what is studied in this paper.

Here, we consider the sensitivity to iDM and visible dark photon decays with NA62 running in a beam-dump mode, in order to go beyond the strong constraints we found based on CHARM and NuCal experiments. We consider runs of $1.3\times 10^{16}$ POT and $10^{18}$ POT impacting on a copper target and assume $N_{\pi^0}/\mathrm{POT}=2.4$ with $N_\eta/N_{\pi^0}$=0.11.

NA62's decay volume is a one-meter radius cylinder with a length of 135 meters located 82 meters downstream from the beam-dump \cite{NA62:2017rwk,Dobrich:2019dxc}. However, for charged particles, since the first spectrometer chamber is located at around 180 meters downstream from the beam-dump and is crucial for particle identification, the fiducial length is a much shorter than 135 m \cite{PrivateCom}. Taking this information into account, we use 75 meters as the fiducial decay length for charged particles. 
Located at the end of the decay volume is a Liquid Krypton Calorimeter (LKr). To study the final state lepton-anti-lepton pairs produced from the decay of exotic particles, we require both leptons to impact the LKr, with a minimum mutual distance of 10 cm to ensure sufficient separation. Both leptons must also be more than 15 cm from the central hole of the LKr \cite{PrivateCom}.
Following ref. \cite{Dobrich:2019dxc}, we assume that the background can be reduced to zero and exclude parameter space for which more than 2.3 events are expected at a 90\% confidence level with an efficiency of $\epsilon=1$, but further study should be conducted to provide a more realistic determination of the background and detection reconstruction efficiency.

As we see in Fig. \ref{fig:1_iDM}, NA62 has slightly worse sensitivity to smaller dark matter masses than other, similar experiments considered, as the sensitivity is heavily suppressed at $m_{\chi_1}=30\,\mathrm{MeV}$ due in large part to the 10 cm separation cut.

\subsection{SeaQuest/DarkQuest Experiment}
\label{sec:seaquest}

SeaQuest is a FNAL fixed-target experiment that studies sea quark distributions inside the nucleon (proton and neutron), utilizing the 120 GeV proton beam from the Main Injector. Recently, it is renamed as "SpinQuest", marking the beginning of project E1039, a physics program that studies sea quark Sivers asymmetry through the Drell-Yan production in the dimuon channel with a polarized target.

Here, we focus on the sensitivity of SeaQuest to iDM in the sizable mass-splitting $\Delta$ regime, primarily motivated by the muon $g-2$ anomaly described in Sec. \ref{sec:iDM_g-2}. 
The prospect of utilizing the SeaQuest experiment to study the dark sector was previously discussed in Refs. \cite{SeaQuest_slides,Gardner:2015wea,Liu:2017ryd,Berlin:2018pwi}. 
Recently, a displaced vertex trigger was installed that allows for the study of dimuons produced through exotic particle decays \cite{dp_trigger}, and a SeaQuest run of $3 \times 10^{16}$ POT was conducted to study long-lived particle decays \cite{PrivateCom}.
In our analysis, we further assume that a proposed Electromagnetic Calorimeter (ECal) upgrade is installed at SeaQuest, repurposing an ECal sector from the recently decommissioned PHENIX experiment at RHIC \cite{Liu:2017ryd,SeaQuest_slides,Berlin:2018pwi}.
This upgrade allows the measurement of dielectron events in SeaQuest to study dark sector-associated events below the current dimuon threshold and improves the background study of dimuon events. This upgrade is crucial for the iDM study, as we study the model through the dielectron decay products from the iDM semi-visible decay.
The experimental proposal for such an upgrade is currently called "DarkQuest" \cite{PrivateCom} and we will uniformly refer to the upgraded setup as SQ/DQ (SeaQuest/DarkQuest).

To achieve a significant improvement on sensitivity to the iDM parameter space given the strong bounds we set by considering CHARM and NuCal data, we include not only SQ/DQ phase I ($1.44\times 10^{18}$ POT), which is expected to begin taking data by the end of 2019 \cite{PrivateCom} but also phase II (with an enhanced $10^{20}$ POT), both assuming a 120 GeV beam impacting a thin nuclear target followed by a magnetized iron beam-dump (FMAG). 
The $10^{20}$ POT improvement of intensity could be a tall order, but can potentially be achieved based on the Fermilab Proton Improvement Plan (PIP) \cite{Shiltsev:2017mle}.

For our analysis, we consider a decay volume beginning after the "FMAG" (a 5\,m-think magnetized iron block that serves as the beam-dump), setting a minimum decay distance 5\,m downstream of the target. 
Note that only $\sim 5 - 10$ $\%$ of the protons interact with the thin target. Most of the proton interactions ($\sim 90 - 95$ $\%$) occur in the first 50\,cm of the FMAG.
The decay region we consider extends from 5\,m to 12\,m downstream from the target, corresponding to a region consisting of SQ/DQ tracking station 1 and the "KMAG," a 3\,m long open-aperture magnet. We require that the charged decay products reach the ECAL behind tracking station 3, 18.5 meters downstream from the target. For SeaQuest, we compared mesons produced with the BMPT distribution and EPOS-LHC \cite{Pierog:2013ria} through the CRMC v1.7 framework \cite{crmc:2019}, and found that the overall event rates differed by less than 20\%. EPOS-LHC predicted 3.3 $\pi^0$ per POT, and 0.11 $\eta$ per $\pi^0$, and these numbers were adopted for the simulation.
The momentum kick of the KMAG is modeled as an instantaneous change in momentum of $0.4\,\mathrm{GeV/c}$ in the $+x$ direction upon exiting the far face of the KMAG, as is discussed in \cite{Berlin:2018pwi}. Should the decay occur inside the KMAG itself, the kick is reduced by a factor of $\Delta z/3\,\mathrm{m}$, where $\Delta z$ is the longitudinal fraction of the KMAG through which the resulting electron-positron pair travels.
In our MC simulation, we found that the kick of particles passing through the KMAG due to its magnetic field is significant in reducing not only the soft SM background but also the electron pair iDM signal from semi-visible decays. 
This effect cancels out the advantage of SQ/DQ being closer to the target, which was thought to be crucial in exploring the iDM muon $g-2$ regime, given a sizable mass-splitting and coupling and thus shorter decay length. 

As demonstrated by our MC, the kick significantly reduces the fraction of electron pairs that survive to station 3, weakening SQ/DQ's sensitivity to iDM decays, especially in the large coupling constant regime.
Therefore, even though the SQ/DQ decay regime and detector are much nearer to the target than any other experiment we considered, it is not much more capable of probing the strong coupling regime nor the large mass regime, when the mass splitting $\Delta$ is sizable. The possibility of shutting down the KMAG magnetic field was discussed in \cite{Berlin:2018pwi} and will be further discussed in the next section.

Again, in this paper, our primary goal is to provide an estimate of SQ/DQ's ability to probe the iDM muon g-2 regime.
However, an involved study of the SQ/DQ background is necessary to determine a realistic sensitivity projection. Here we would like to discuss a potential concern with the SQ/DQ background, especially during the $10^{20}$ POT run. 
As estimated in \cite{Berlin:2018pwi}, assuming the misidentification of pions from neutral kaon decay is on the level of one percent, the level of background from $K^0_L\rightarrow \pi^{\pm} e^{\mp} \nu$ is on the level of $\sim$ 10 events in the $10^{18}$ POT run, though pointing information may help further reduce the background. The background would likely scale up to $\sim$ $10^3$ events in the $10^{20}$ POT run (not yet including the estimation of other SM backgrounds), and could pose a significant challenge for the new physics search. Thus, we believe a detailed reconsideration of the SQ/DQ setup would be beneficial in the preparation of the $10^{20}$ POT run, and we conducted rough estimates in support of this objective as detailed in the next section.

\subsection{LongQuest: A Complete Retool of SeaQuest/SpinQuest Experiment}
\label{sec:LongQuest}

In anticipation of a potential $10^{20}$ POT run, we take a second look at the SQ/DQ experimental configuration and consider scenarios that could fulfill its full new-physics potential beyond the capability of the DarkQuest upgrade. We consider LongQuest, a dedicated and complete retool of SQ/DQ to search for long-lived particles with improved and additional decay detectors, extra long based-line detectors, and additional beam-dump and tracking detectors in front of the current target, as discussed below:

\subsubsection{LongQuest-I: Getting RICH, KMAG Reduction, $\&$ Extended Fiducial Volume}

There are several installations one can consider in order to improve the SQ/DQ detector, beyond the previously proposed installation of an ECal. 

First, as discussed in the previous section, neutral kaon decay may result in background events on the level of $\sim$ $10^3$ events during the $10^{20}$ POT run. In order to reduce this substantial background, one could hope to achieve a better separation between pions and electrons through the implementation of a Ring-Imaging Cherenkov (RICH) detector or a Hadron Blind Detector (HBD) \cite{Akiba:2000up,HBD}.
RICH detectors are commonly used for particle identification (PID) for electrons and hadron separation, utilizing the fact that only electrons can produce Cerenkov light, while the hadrons generally can not, in these dilute gas detectors. More specifically, one can install the PHENIX RICH detector \cite{Akiba:2000up} between station 2 and 3 to differentiate between pions and electrons. A detailed consideration of both possibilities would be mapped out in \cite{LongQuest_Proposal}.

Secondly, in light of KMAG's suppression of new physics events with charged low-energy final states \cite{Berlin:2018pwi}, we consider turning off or tuning the KMAG to improve the sensitivity for the models like iDM that involve 3-body decays and lower-energy visible decay products.
However, given that the strong magnetic field is used to filter out the soft SM background, one would need to compensate for the loss of this ability with additional detector installations (the aforementioned RICH detector or HBD additions) or a stronger analysis cut (e.g., a stronger energy-threshold cut) to reduce the additional soft SM background.
The effects of different energy-threshold cuts are discussed in Sec. \ref{sec:discussion_iDM}.

Finally, one can consider extending the fiducial volume, as a large decay volume is beneficial for the dark sector particle search. 
One can try to extend the SQ/DQ displaced dark sector dielectron or dimuon decay volume from the previously discussed 5 m -- 12 m to 5 m -- 18 m, almost saturating the maximum potential decay length for iDM to which SeaQuest is sensitive.
One would still install the ECAL between stations 3 and 4 or place the ECAL behind station 4. Also, more detector stations and a magnetic field weaker than the current KMAG to split the lepton pairs can be added in front of station 3 (18.5 meters downstream of the target). 
It is worth noting that even the 18-meter benchmark is not necessarily the maximum possible length for the decay volume. One can consider improving or replacing station 3 (and all other stations) with better technology and potentially extending the decay volume further. 
It is, however, evident that doubling the length of the decay volume would at most double the event rate when the decay length of the particle is much longer than the length of the detector. The background in an extended decay volume could also be larger.
It is still interesting to consider this possibility given that LongQuest provides one of the very few near-future and low-cost opportunities\footnote{Comparing to DUNE and SHiP \cite{Alekhin:2015byh,Abi:2018dnh}} to further extend the sensitivity reach to lower couplings in the MeV to GeV regime for the minimal dark photon model, as shown in Fig. \ref{fig:dp_update}.

In this paper, we conduct a phenomenological estimation assuming that the background can be understood and reduced in the $10^{20}$ POT run. 
We provide two curves for this $10^{20}$ POT run.
One labeled "SQ/DQ (1e20)" that assumes a total of $10^3$ expected background events from $K^0_L$ and other SM sources.
We consider another curve, "LongQuest-I," assuming the expected background is reduced by 90 percent based on the installations of the new detectors and that the KMAG is shut off (still 5 m -- 12 m fiducial decay length).
We {\it do not} consider an extended fiducial volume for LongQuest-I because the change could introduce more unexpected background events.
Note that these are rough estimates of the background and could certainly be over or under a more realistic number, but we use this as a demonstration and motivation for further study.
In Fig. \ref{fig:iDM_g-2}, we show the estimation of the sensitivity of both SQ/DQ (1e20) and LongQuest-I sensitivity. 
The reduced background certainly helps the sensitivity in the weaker coupling regime, and we also find that the reduction of the KMAG magnetic field improves the ability of LongQuest-I to probe of the high mass and high coupling regime for iDM, along with the reduced background. The enhanced sensitivity significantly helps the LongQuest-I study of the muon $g-2$ motivated region.
Given the removal of KMAG magnetic field, we impose a conservative 3 GeV analysis cut for our analysis of the LongQuest-III sensitivity for minimal dark photon as a way to reduce soft SM background. For the iDM muon $g-2$ regime, we conduct analyses with both 3 GeV and 0 GeV cuts to show the effect of the energy cut. A realistic analysis cut should lay between 0 and 3 GeV.

The technology and experimental configurations developed, for example, in \cite{Feng:2017uoz,Gligorov:2017nwh} can be applied here to improve the dark sector search further.
Furthermore, the addition of a RICH detector or HBD would also allow different analyses with sensitivity to additional dark sector scenarios, allowing LongQuest-I to be a multi-purpose experiment.

\subsubsection{LongQuest-II: Long Baseline Detectors}

Previous studies of the SQ/DQ dark sector search \cite{SeaQuest_slides,Gardner:2015wea,Berlin:2018pwi} focused on the study of displaced lepton signature that can be seen by the ECal-upgraded SeaQuest/SpinQuest detector. Here, we propose an exciting new opportunity to utilize the same 120 GeV main injector beam but place detectors further downstream of the present location of the primary detector.

We note that there are empty spaces behind the 10-meter iron block behind the SQ/DQ station 4 which could accommodate additional detectors for a search for long-lived dark sector particles. In particular, there is another ECAL from the Phenix experiment that could be installed, and this space is also an ideal place to host a dedicated search for millicharged particles (see the FerMINI \cite{Kelly:2018brz} experimental concept.)
% or even neutrino physics studies

Given that a 10-meter iron block further reduces the background, the detector can conduct a extremely-low background study, comparing to the SQ/DQ setup.
We consider the installation of this new detector (or detectors) as "LongQuest-II" and the design and physics cases will also be detailed in \cite{LongQuest_Proposal}. 

\subsubsection{Prompt Dimuon Analysis and LongQuest-III}

Finally, we comment on a specific analysis and a new experimental configuration that can further improve the minimal dark photon study. As discussed in \cite{SeaQuest_slides}, it is interesting to search for the dimuon signature from prompt dark photon $A^{'}$ decay, where they were produced in the beam-dump but travel less than five meters and decay within the FMAG. The muons could penetrate through FMAG and register at detector stations 1 -- 4.
The signature would allow one to access a dark photon with a very short decay length and thus large coupling. 
However, the SM background is very large since a large number of dimuons would be produced promptly through SM processes. The search is thus very involved and challenging. However, since the SM dimuon signature is one of the primary goals for SeaQuest \cite{SeaQuest_slides}, the dedicated simulation of the SM events is an active ongoing effort.

One idea that under consideration \cite{PrivateCom} is that one could add an additional beam-dump in front of the current target and FMAG, and add a fast tracking detector in between the new beam-dump and FMAG to help signature analysis and background reduction. This potential upgrade which we refer to as LongQuest-III could be crucial for the prompt dimuon search and will also be discussed in \cite{LongQuest_Proposal}. \\

\section{Results and Discussions} \label{sec:discussion}
\subsection{iDM Thermal Target and $(g-2)_\mu$}
\label{sec:discussion_iDM}

%\onecolumngrid
%!h
%\lipsum
\begin{figure*}[]
    \centering
        \subfloat[iDM: $\Delta=0.05,$ $\alpha_D$=0.5.]{\includegraphics[width=0.49\textwidth]{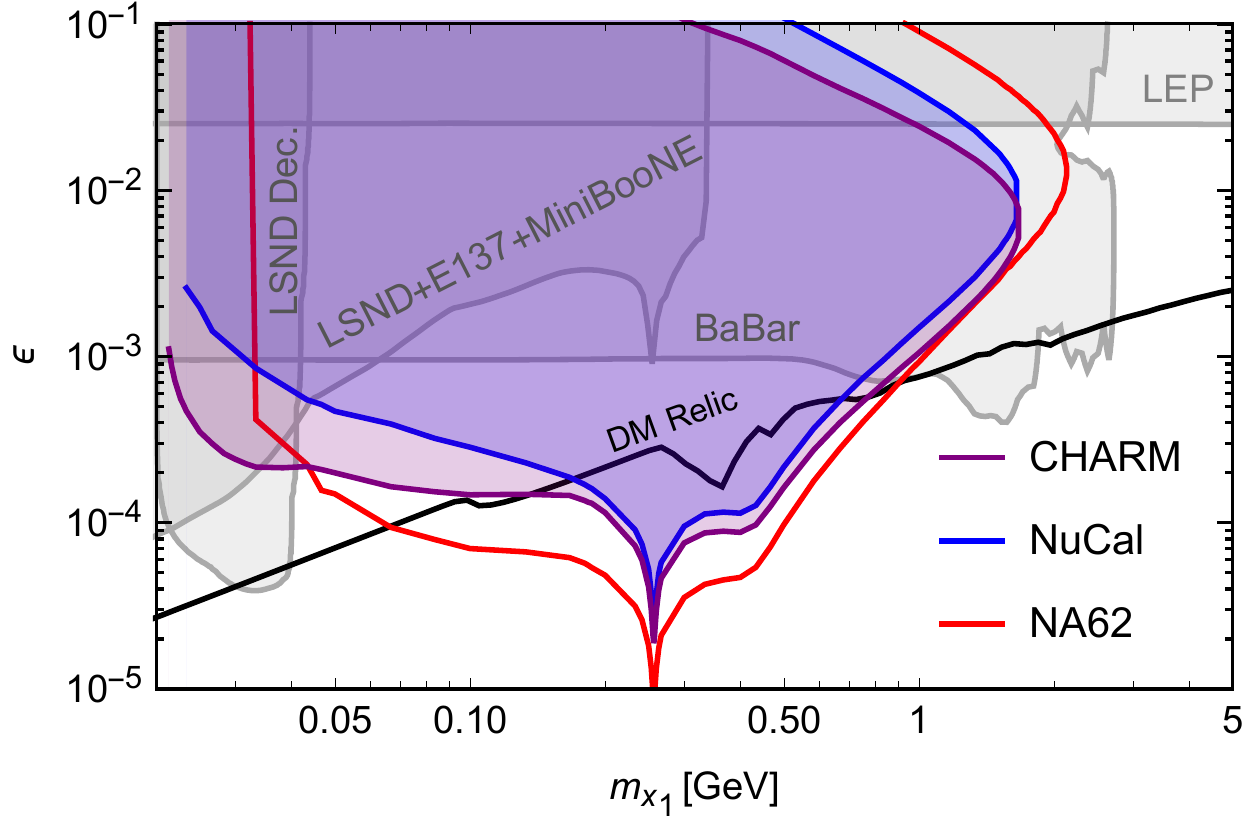}} 
        \label{fig:1a}
        \subfloat[iDM Thermal Target: $\Delta=0.05,$ $\epsilon=\epsilon_{\rm relic}$.]{\includegraphics[width=0.49\textwidth]{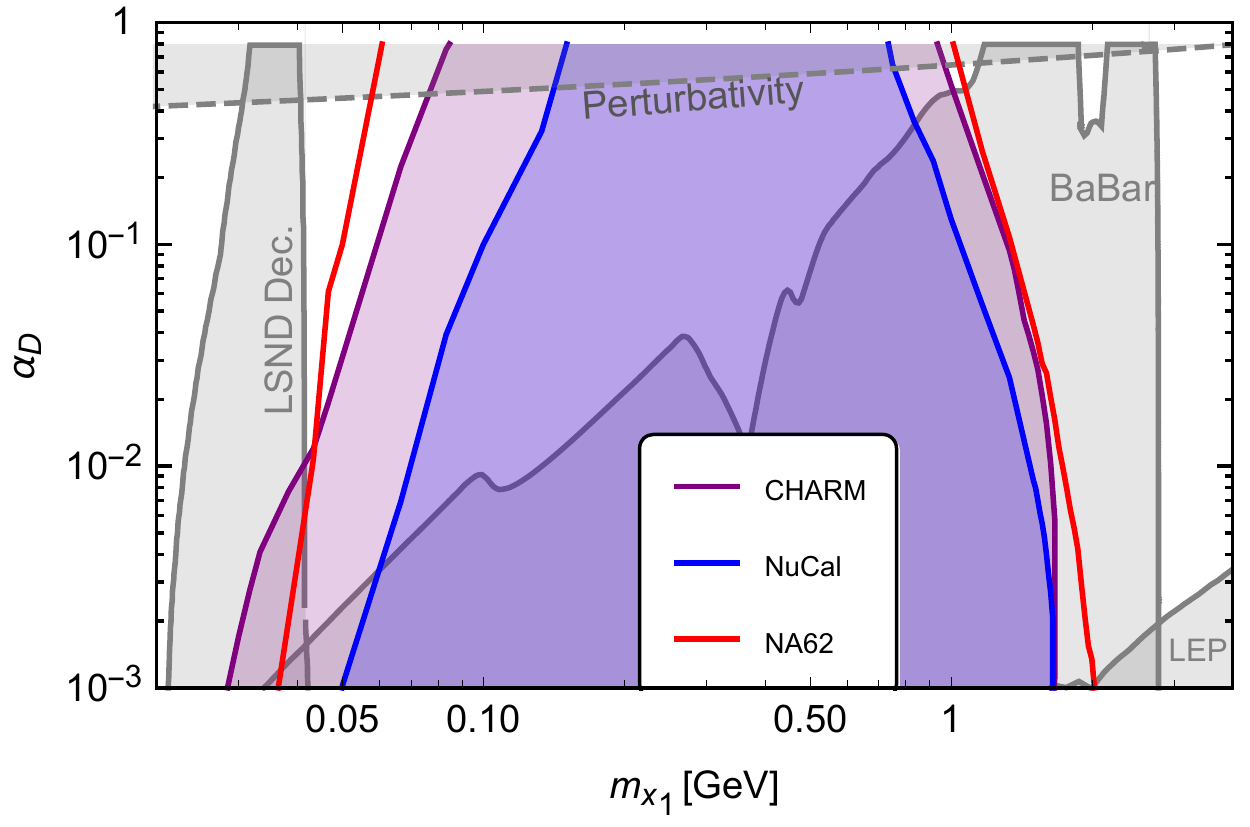}}
        \label{fig:1b}  
        \subfloat[iDM: $\Delta=0.1,$ $\alpha_D$=0.1.]{\includegraphics[width=0.49\textwidth]{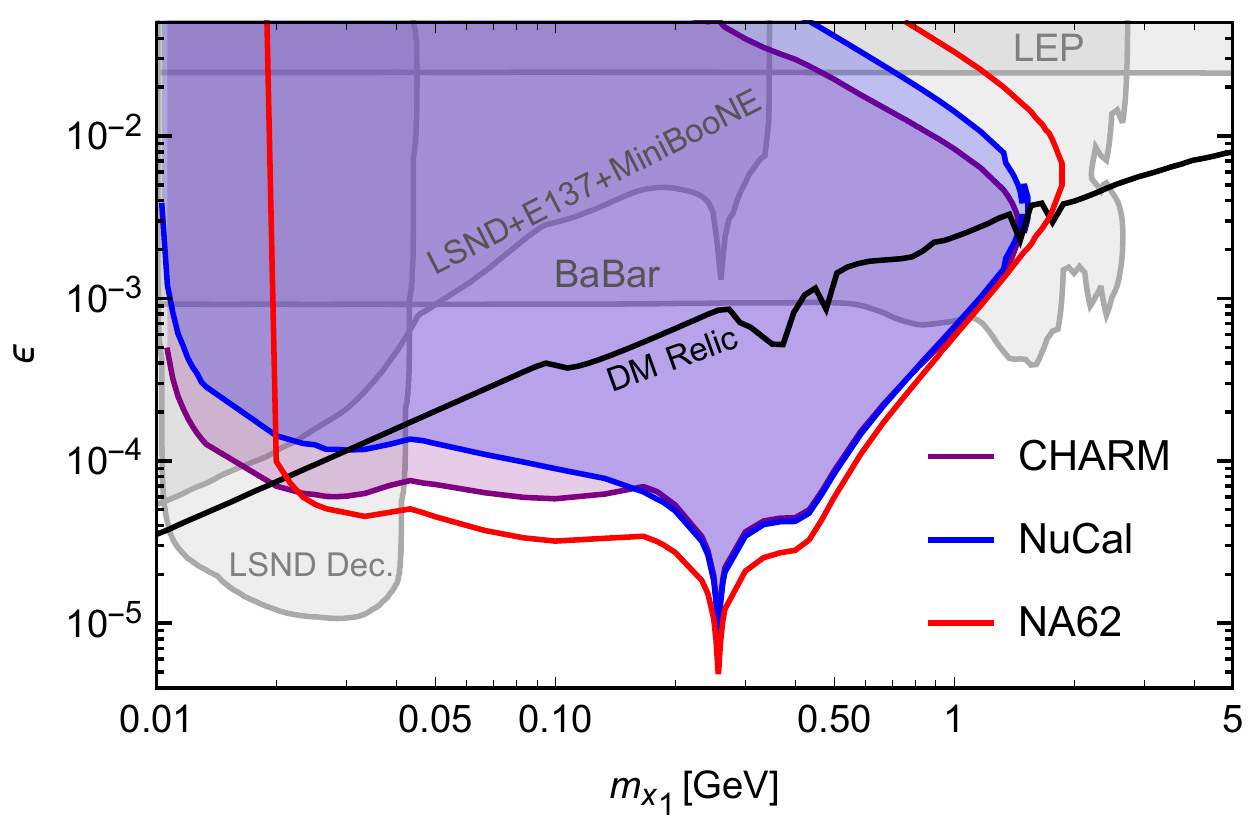}}
        \label{fig:1c}    
        \subfloat[iDM Thermal Target: iDM $\Delta=0.1,$ $\epsilon=\epsilon_{\rm relic}$.]{\includegraphics[width=0.49\textwidth]{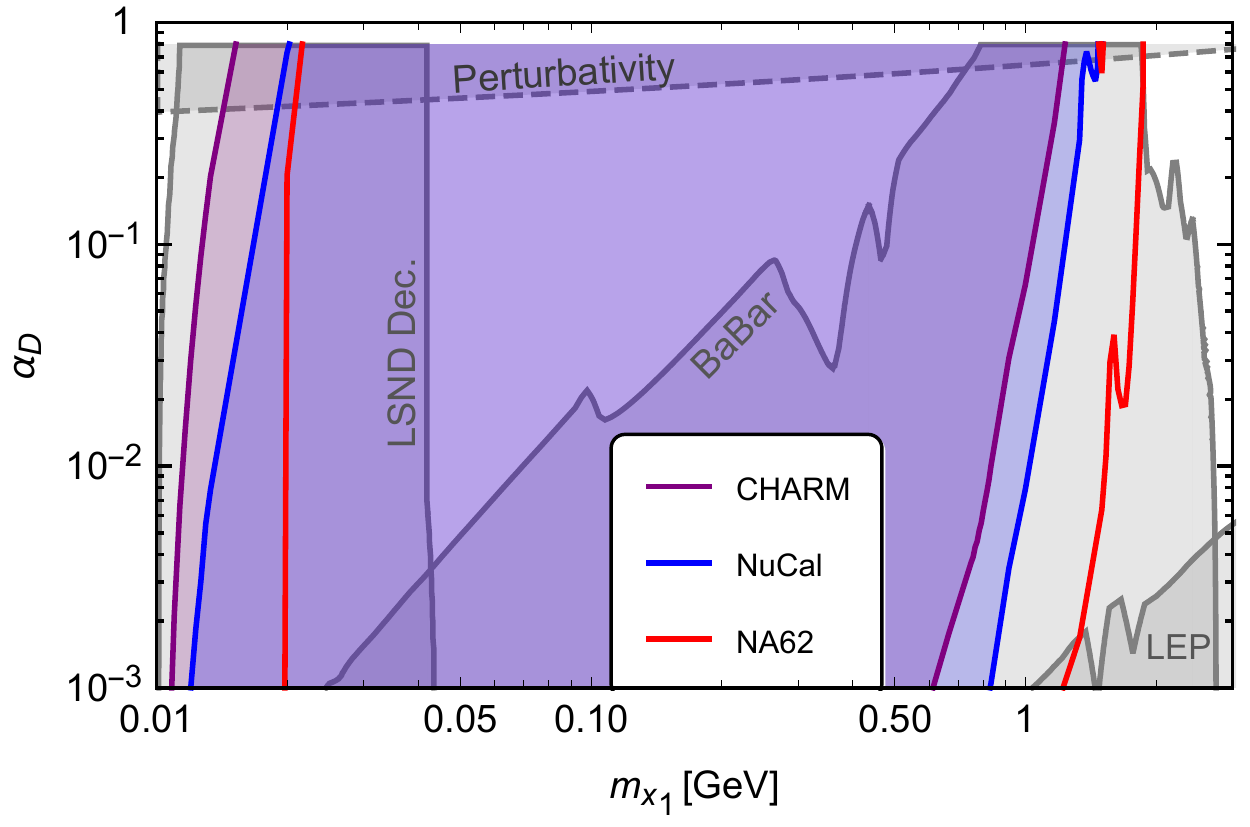}}
        \label{fig:1d}            
        \subfloat[ \label{fig:full}  
        Compilation of relevant constraints and sensitivity projections for iDM with $\alpha_D=0.1$ and $\Delta=0.1.$]{\includegraphics[width=0.85\textwidth]{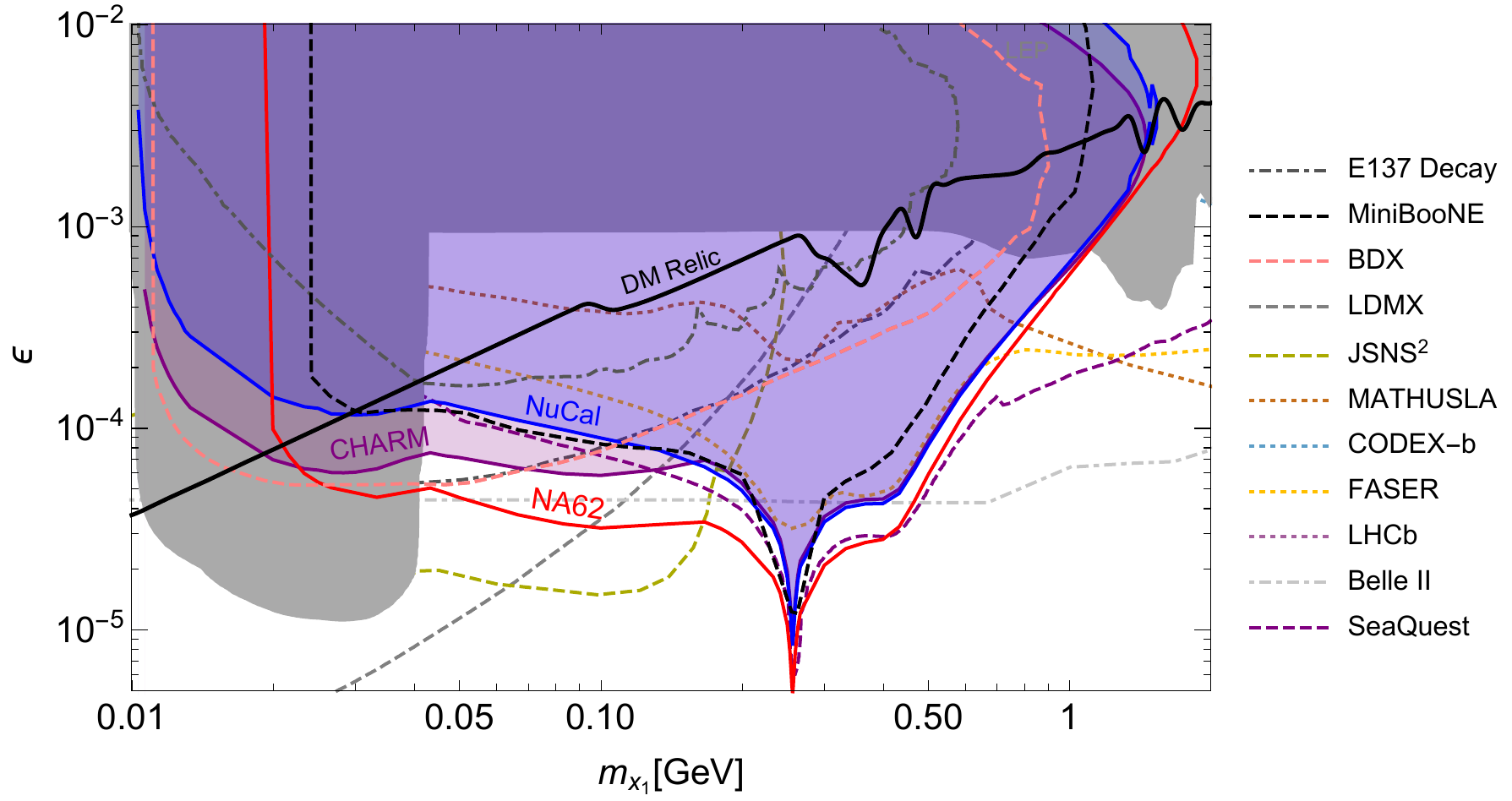}}
\caption{
We show new constraints on iDM based on the data of CHARM ({\purple purple}) and NuCal ({\blue blue}), and the projected sensitivity of the NA62 beam-dump run ({\red red}). The gray shaded regions are previously existing constraints, excluding E137 decay (see Sec. \ref{sec:discussion_iDM}). 
In (e), we include the potential E137 decay constraint \cite{Bjorken:1988as,Izaguirre:2017bqb,Berlin:2018pwi} along with projections from MiniBooNE \cite{Izaguirre:2017bqb}, BDX \cite{Izaguirre:2017bqb}, LDMX \cite{Berlin:2018bsc,Berlin:2018pwi,Akesson:2018vlm}, $\rm JSNS^2$ \cite{Jordan:2018gcd}, MATHUSLA \cite{Berlin:2018jbm,Chou:2016lxi}, CODEX-b \cite{Berlin:2018jbm,Gligorov:2017nwh}, FASER \cite{Berlin:2018jbm,Feng:2017uoz}, LHCb \cite{Ilten:2016tkc,Aaij:2017rft,Pierce:2017taw}, Belle-II \cite{Kou:2018nap}, SeaQuest \cite{Berlin:2018pwi}. Other probes outside of this parameter space (see, e.g., \cite{Curtin:2014cca,Izaguirre:2015zva,Liu:2018wte}) are not shown. One can see E137, MiniBooNE and BDX projections are already covered by CHARM and NuCal.}
 \label{fig:1_iDM}     
\end{figure*}
%%%%%%%%%%%%%%%%%%%%%%%%%%%

\begin{figure*}[]
    \centering
        \subfloat[iDM: $\Delta=0.4,$ $\alpha_D$=0.1. With muon $g-2$ and DM regimes.]{\includegraphics[width=0.49\textwidth]{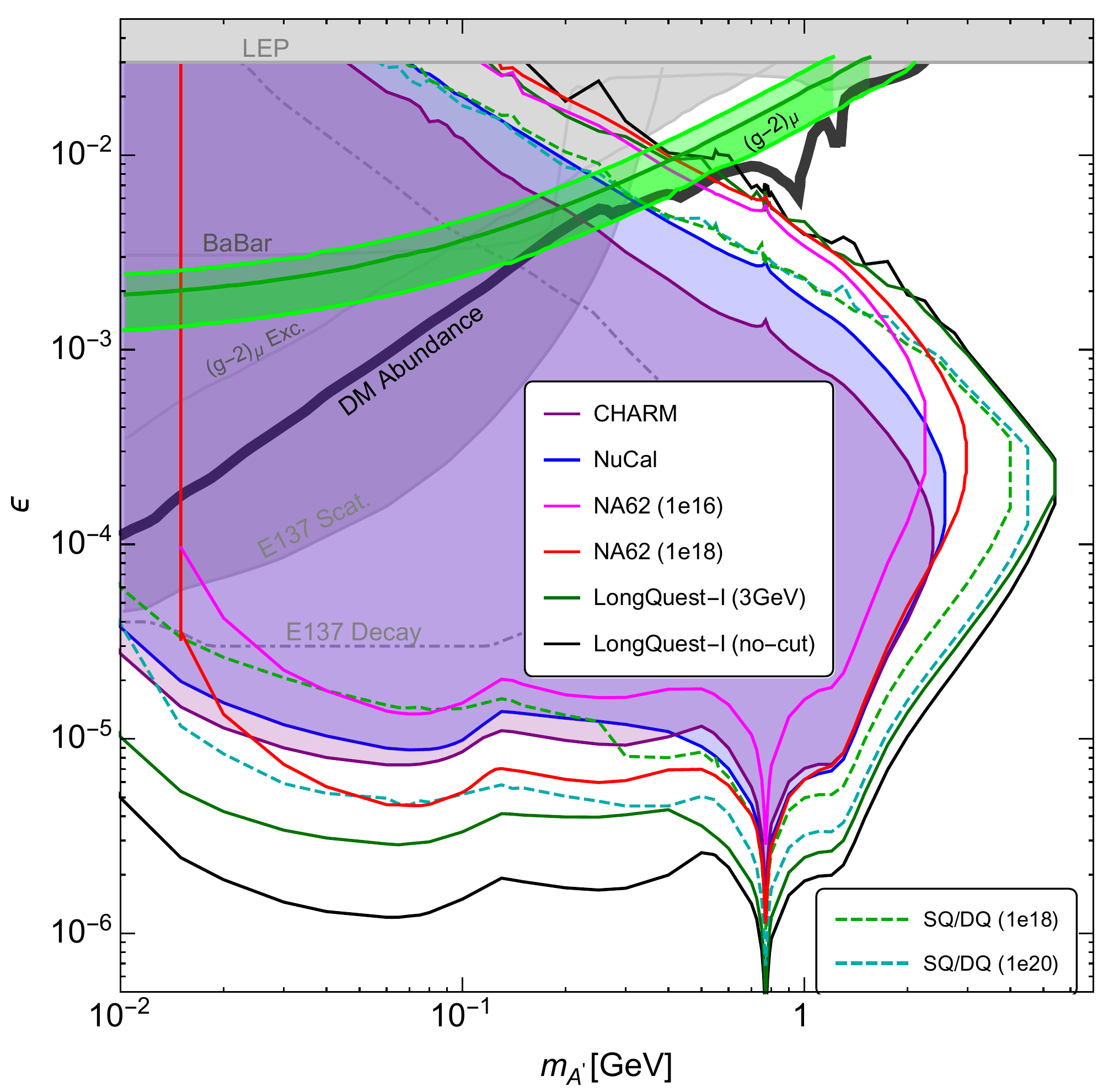}} 
        %\label{fig:1a}
        \subfloat[iDM Muon $g-2$ Target: $\Delta=0.4,$ $\epsilon=\epsilon_{(g-2)_\mu}$.]{\includegraphics[width=0.49\textwidth]{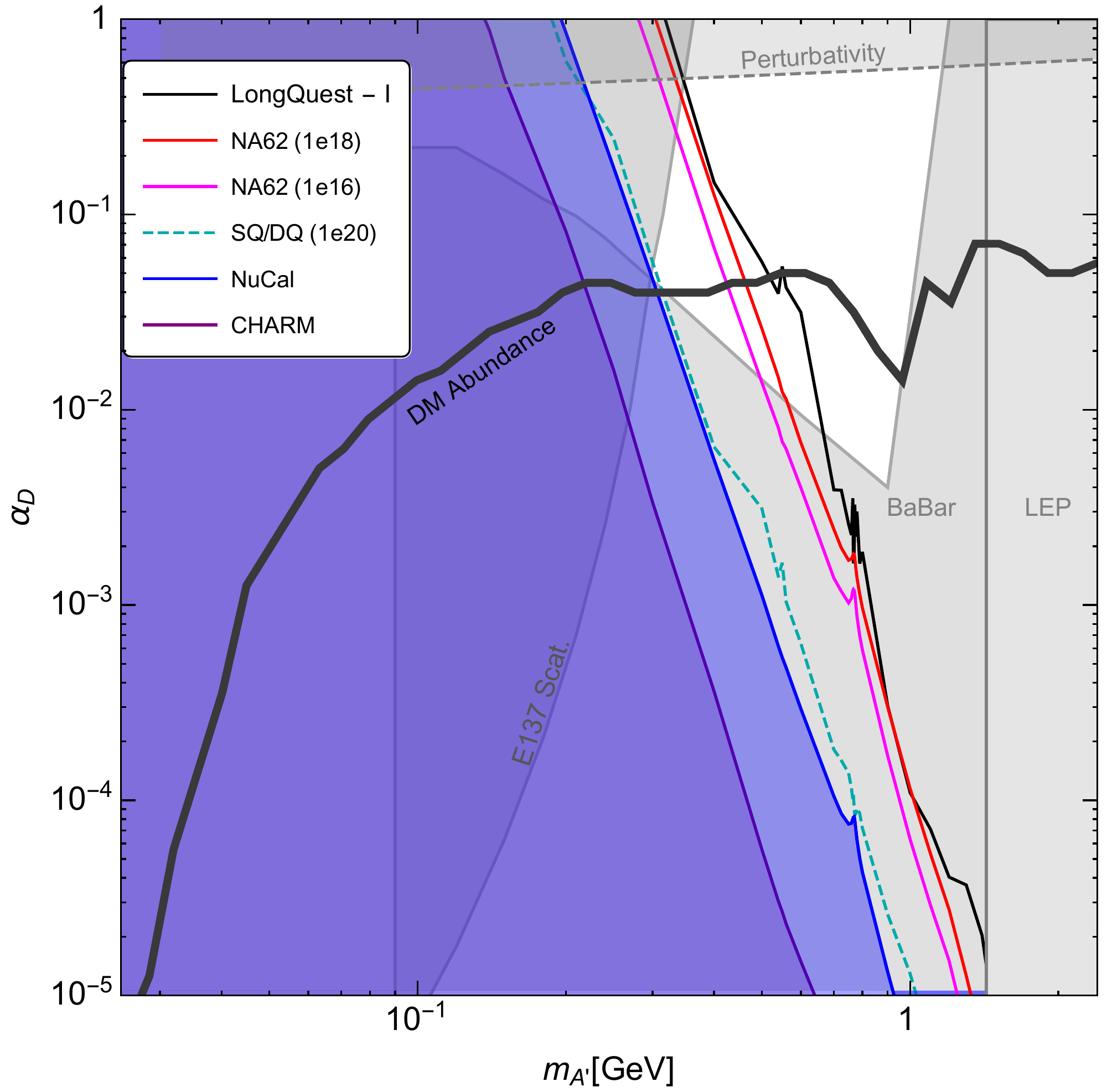}}    
  \caption{This plot shows constraints and sensitivity projections for iDM within the muon $g-2$ motivated regime. The muon $g-2$ favored regime is the light-green band while the thick-black curve is again the parameter contour yielding the correct DM relic abundance.
We considered the bounds from CHARM ({\purple purple}) and NuCal ({\blue blue}), and projections from NA62 ($10^{18}$ POT: {\red red}, $1.3\times 10^{16}$ POT: {\magenta magenta}), SQ/DQ ($10^{20}$ POT: {\cyan~dashed-cyan}, 1.4$\times 10^{18}$ POT: {\pinegreen~dashed-pine green}), and LongQuest-I (3-GeV cut:{\green~darker green}, no-cut: {\bf black}). In (b), we only plot LongQuest-I no cut curve because it is basically identical as the 3-GeV cut result. The gray region is the previously existing constraints.}
\label{fig:iDM_g-2}
\end{figure*}
%%%%%%%%%%%%%%%%%%%%%%%%%%%%

\begin{figure*}[]
    \centering
        \subfloat[\label{fig:dp_update_a}
        Updates on dark photon bounds and the NA62 projection.]{\includegraphics[width=0.49\textwidth]{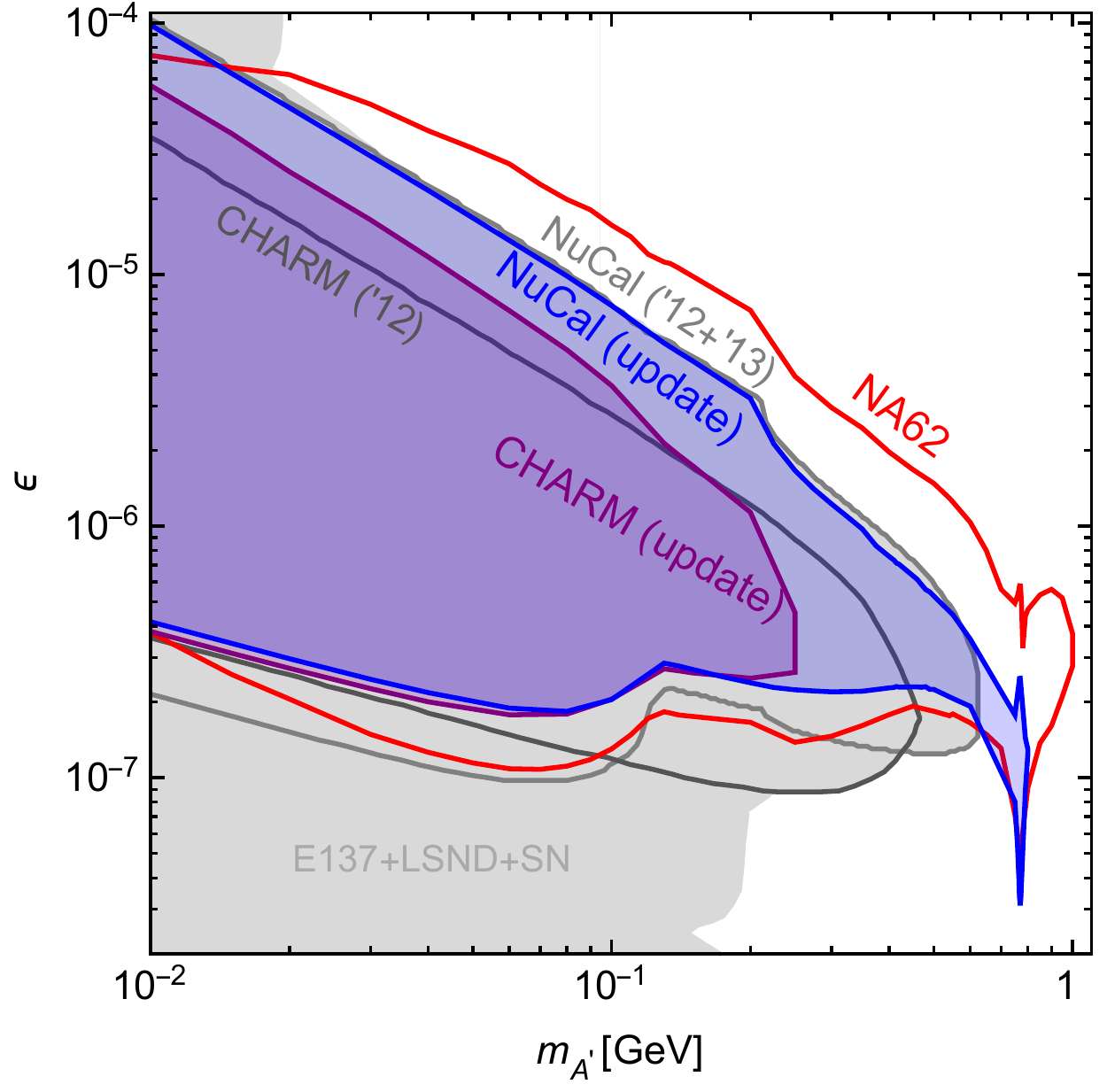}} 
        \subfloat[\label{fig:dp_update_b}
        Compilation of projections and constraints on dark photon.]{\includegraphics[width=0.49\textwidth]{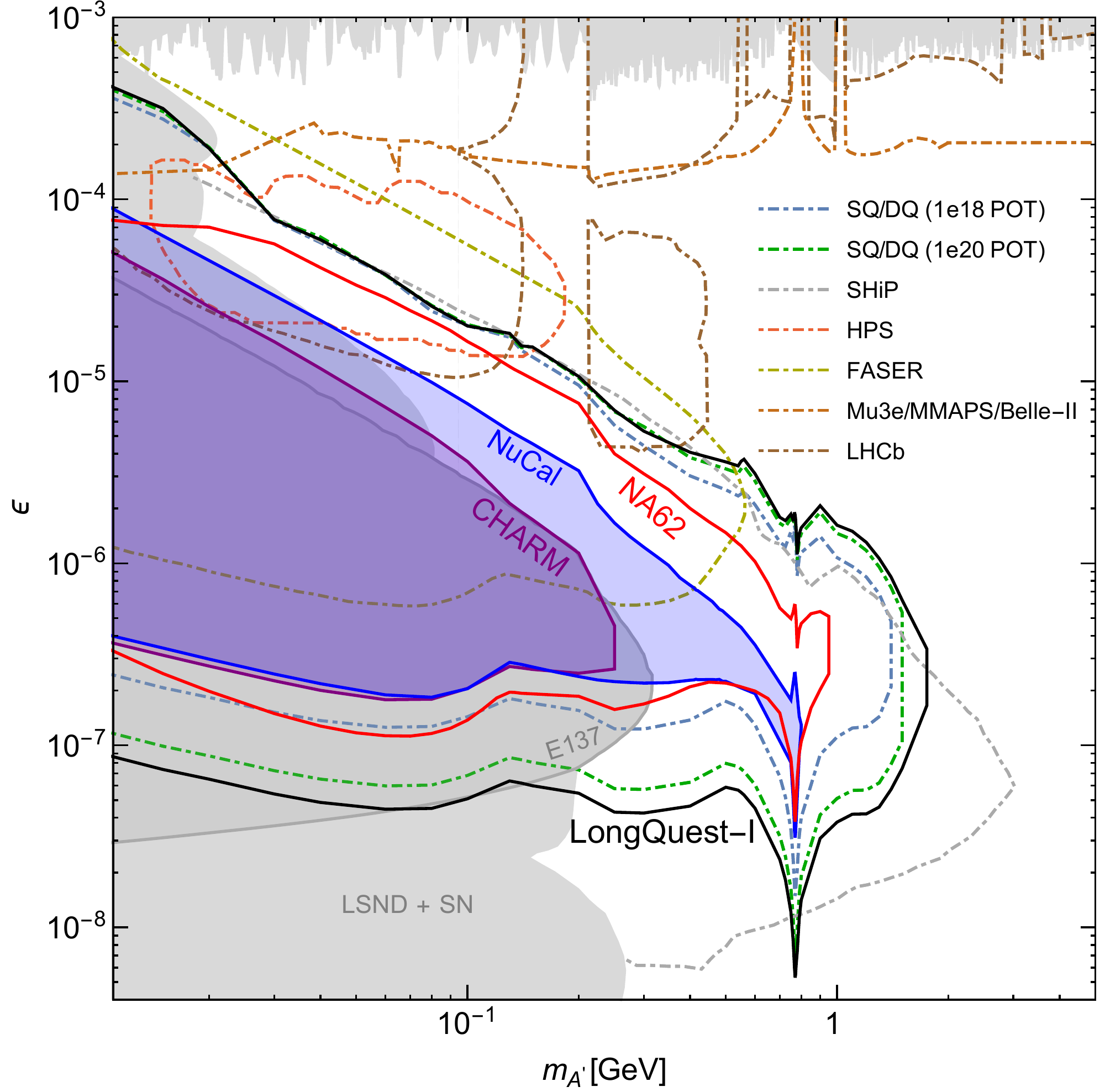}}    
  \caption{We show updates on the kinetically-mixed visibly decaying dark photon constraints and projections. In (a), gray contours  are the previous bounds set based on analyses of NuCal \cite{Blumlein:2011mv,Blumlein:2013cua} and CHARM \cite{Gninenko:2012eq} experiments. In (b), projections for future experiments are shown in dot-dashed curves with color and also labeled in the plot.
In both (a) and (b), our updated bounds on CHARM ({\purple purple}), NuCal ({\blue blue}), and new projections of NA62 ({\red red}) and LongQuest-I ({\bf black}) are shown.
Note that we do not find any fundamental disagreement with the pioneering works on dark photon constraints from CHARM and NuCal \cite{Blumlein:2011mv,Gninenko:2012eq,Blumlein:2013cua}. We add relevant production channels and have a conservative assumption of meson production rates. Readers can take our CHARM and NuCal constraints as conservative limits.}
  \label{fig:dp_update}
\end{figure*}

%%%%%%%%%%%%%%%%%%%%%%%

\twocolumngrid

The constraints and sensitivity projections for iDM with small $\Delta=0.05,0.1$ are shown in Fig. \ref{fig:1_iDM} and the parameter space to explain muon $g-2$ anomaly in Fig. \ref{fig:iDM_g-2}. 
All previously considered experimental constraints in this parameter space are included and labeled in the plots \cite{Bjorken:1988as,Auerbach:2001wg,Chung:2003zr,Aubert:2008as,Hook:2010tw,deNiverville:2011it,Batell:2014mga,Curtin:2014cca,Lees:2017lec,Aguilar-Arevalo:2017mqx,Berlin:2018pwi}, with the exception of the potential constraints from the recast of the search for neutral exotic particle decays (e.g. axion-like particles and photinos) in the E137 experiment \cite{Bjorken:1988as}. 

The recast of the E137 analysis could potentially provide bounds on iDM in these regimes. However, there are two challenges in generating robust constraints from this E137 analysis \cite{Bjorken:1988as}.
First, various energy-threshold cuts (1-3 GeV) were applied to different stages of the data taking and analysis in \cite{Bjorken:1988as}\footnote{As indicated in \cite{Bjorken:1988as}, the data were taken with a 1 GeV threshold cut and a 2 GeV analysis cut in the main text. However, in the Appendix of \cite{Bjorken:1988as}, it was indicated that there is a detector-level cut on the energy spectrum at "2 or 3 GeV", thus creating difficulties for a recast.}, and it is not clear which cut to use to derive a robust bound for iDM, as also discussed in \cite{Berlin:2018pwi}.
Secondly, the data were analyzed through a visual display, which may not meet the standard of today's experiments.
As discussed, the iDM 3-body decay signature is very sensitive to the energy threshold, unlike the original search for axion-like particles in \cite{Bjorken:1988as}.
If one chooses the most conservative 3 GeV threshold, the E137 bounds for iDM will be worse than the combination of other previously existing bounds. 

Here, we do not present the potential E137 decay bound the same way as the other existing constraints, because a dedicated effort (and potentially access to the original data set) is needed to determine robust constraints properly. 
Instead, we show the potential E137 constraints estimated from \cite{Berlin:2018pwi,Mohlabeng:2019vrz} as gray dot-dashed curves in both Fig. \ref{fig:1_iDM} and Fig. \ref{fig:iDM_g-2}. One can see that the E137 curves are already covered by CHARM and NuCal constraints. A quick estimation shows that the E137 decay bounds from \cite{Bjorken:1988as} are always weaker than the CHARM and NuCal constraints we derive, regardless of the cuts.  

As shown in Fig. \ref{fig:1_iDM}, CHARM and NuCal provide strong constraints on iDM with small mass splittings, almost excluding all of the regimes that predict the correct DM relic abundance. NA62, with its projected sensitivity shown by a dotted purple line, can further improve the exploration of these iDM scenarios.

Other future probes of iDM \cite{Berlin:2018jbm}, including LHC (ALTAS and CMS) \cite{Izaguirre:2015zva,Liu:2018wte}, LHC EWPT test \cite{Curtin:2014cca}, LHCb \cite{Ilten:2016tkc,Aaij:2017rft,Pierce:2017taw},
SeaQuest \cite{Berlin:2018pwi},
Belle-II \cite{Kou:2018nap},
CODEX-b \cite{Gligorov:2017nwh}
FASER \cite{Feng:2017uoz}, MATHUSLA \cite{Chou:2016lxi}, Babar \cite{Izaguirre:2015zva},
MiniBooNE \cite{Izaguirre:2017bqb},
$\rm JSNS^2$ \cite{Jordan:2018gcd},
BDX \cite{Izaguirre:2017bqb}, and
LDMX \cite{Berlin:2018bsc,Berlin:2018pwi,Akesson:2018vlm}
which can also improve the limits on iDM with small mass splittings. The ones that are relevant for the 10 MeV to GeV regime are also included in Fig. \ref{fig:full}.

For the large mass splitting iDM parameter space motivated by the muon $g-2$ anomaly, the proton fixed-target experiments also provide strong new constraints, as shown in Fig. \ref{fig:iDM_g-2}.
In order to resolve the muon $g-2$ discrepancy, there must be large kinetic mixing between the dark and SM photons, and the dark photon must have a mass greater than $\sim$ 300 MeV to escape existing constraints. The CHARM and NuCal analyses poorly constrain this still-open muon $g-2$ favored parameter space because the combination of large mass and strong coupling to the SM leads the $\chi_2$ particles to decay before they reach the fiducial decay regions.
In principle, SQ/DQ should provide the best sensitivity probes in this regime, given that it has the shortest distance between the target and the fiducial decay region among all the experiments we consider. However, we found that the sensitivity to this regime is only comparable to that of CHARM and NuCal. The reason for this is understood: the strong magnetic field in the KMAG suppresses new physics events by kicking the visible charged products out of the decay volume \cite{Berlin:2018pwi}. This effect is particularly significant for iDM because the lepton pairs from $\chi_2$ decays are soft, given that the $\chi_1$ takes a large fraction of the $\chi_2$ energy. 
The signal suppression from the KMAG effectively cancels out the benefit of a shorter baseline of SQ/DQ. 
In Fig. \ref{fig:iDM_g-2}, we show the first study of the iDM muon $g-2$ regime in SQ/DQ assuming a sizable mass splitting $\Delta$ = 0.4. 
We place a contour on greater than ten events for the sensitivity of SQ/DQ $10^{18}$ POT run. 
For the SQ/DQ $10^{20}$ POT run, we assume $10^3$ background events (a rough estimation from long-lived kaon decays) to set a sensitivity projection.
One can see that SQ/DQ does improve the sensitivity to smaller couplings with a $10^{20}$ POT run. 
However, as can be seen in this figure, the sensitivity to iDM in the $(g-2)_\mu$ valid regime is not improved by SQ/DQ phase I and II.
We then show that the proposed LongQuest-I upgrades with reduced background and no KMAG magnetic field would help explore the iDM muon g-2 regime in Fig. \ref{fig:iDM_g-2}.
For the LongQuest-I analysis, we assume the background is reduced to 10 percent of the SQ/DQ background. Given the removal of KMAG and the potential soft SM background, we consider the sensitivity of a 3-GeV energy threshold cut and a projection with no energy threshold cut. A realistic analysis may need to apply an energy cut between these choices.

Below roughly 10\,MeV, the iDM model as well as other thermal light dark matter are subject to $N_{\rm eff}$ bound \cite{Boehm:2013jpa,Green:2019glg}, and the proton fixed target probes become suppressed by the limited phase space available to the $\chi_1 l^+ l^- $ states. We, therefore, choose to focus on iDM masses larger than 10\,MeV.

\subsection{Minimal Dark Photon} 
\label{sec:discussion_dp}

We consider a dark photon with a mass between MeV and 5 GeV. Many experiments have been proposed to explore this region of the dark photon parameter space, in large part because the dark photon can serve as the mediator between dark matter and the SM to set the relic abundance, or explain experimental anomalies, e.g., the aforementioned muon $g-2$ anomaly, and LSND/MiniBooNE anomalies \cite{Bertuzzo:2018itn}.
The open parameter space of $\epsilon \sim 10^{-8} - 10^{-3}$ avoids collider constraints \cite{Aubert:2009cp,Curtin:2013fra,Lees:2014xha,Ablikim:2017aab,Aaij:2017rft,Archilli:2011zc,KLOE:2016lwm}, beam-dump bounds \cite{Bergsma:1985is,Bergsma:1985qz,Bernardi:1985ny,Konaka:1986cb,Riordan:1987aw,Bjorken:1988as,Bross:1989mp,Davier:1989wz,Blumlein:1990ay,Blumlein:1991xh,Athanassopoulos:1997er,Astier:2001ck,Bjorken:2009mm,Essig:2010gu,Williams:2011qb,Gninenko:2011uv,Abrahamyan:2011gv,Blumlein:2011mv,Merkel:2011ze,Gninenko:2012eq,Blumlein:2013cua,Andreas:2012mt,Merkel:2014avp}, rare-meson decays~\cite{Bernardi:1985ny,MeijerDrees:1992kd,Archilli:2011zc,Gninenko:2011uv,Babusci:2012cr,Adlarson:2013eza,Agakishiev:2013fwl,Adare:2014mgk,Batley:2015lha,KLOE:2016lwm},
and can potentially be explored by various searches in the near future \cite{Essig:2010xa,Freytsis:2009bh,Balewski:2013oza,Wojtsekhowski:2012zq,Beranek:2013yqa,Raggi:2014zpa,Echenard:2014lma,Battaglieri:2014hga,Alekhin:2015byh,Gardner:2015wea,Ilten:2015hya,Curtin:2014cca,He:2017ord,Kozaczuk:2017per,Ilten:2016tkc,Feng:2017uoz,Alexander:2017rfd}.

We update the dark photon bounds from CHARM and NuCal, taking into account additional relevant production channels (production of dark photons from $\eta$ meson decays for NuCal and proton bremsstrahlung for CHARM).
The results are shown in Fig. \ref{fig:dp_update_a}. Interestingly, the bound does not get stronger for most parameter space in our consideration, except for the high mass regime, even though we take into account new relevant production channels. The strength of the bounds in the meson-decay dominated regime ($m_{A^\prime} \ll m_\eta$) are highly dependent on the estimate of the overall meson production rate, and ours are conservative (see Ref. \cite{Dobrich:2019dxc} for some comparisons between different methods of estimating $\pi^0$ production). Our constraint does not show the same mass reach as CHARM largely due to this fact, as we predict an event rate several times smaller than previous analyses.

At larger masses, the primary difference comes from our handling of the proton bremsstrahlung. We consider a timelike form-factor \cite{Faessler:2009tn,deNiverville:2016rqh} which includes a strong resonant enhancement from mixing with the $\rho$ and $\omega$ mesons, visible in the constraint plots as a sharp peak extending the NuCal contours to smaller values of the coupling. The NuCal contour is also the first constraint of this kind with a dark photon mass reach close to 1 GeV, as shown in Fig. \ref{fig:dp_update}. 
One would expect the constraint from CHARM to be enhanced in the same fashion. However, the constraint from CHARM is weakened in part by its off-axis location, given that proton bremsstrahlung production is highly collimated with the beam. 

We also conducted a study of NA62's sensitivity based on the information and selection cuts from \cite{Dobrich:2019dxc,PrivateCom}. We find that it could explore stronger couplings due in large part to the higher energy beam and the same $\rho$ peak exhibited by the NuCal curve appears at larger masses.
In Fig. \ref{fig:dp_update_b}, we compare the updated NuCal and CHARM bounds as well as the NA62 sensitivity reach to previously established constraints and the projection of other future proposals. 

Finally, we make projections for the sensitivity of SQ/DQ and LongQuest-I with a simplified detector simulation. The contours are somewhat bumpy in the large coupling regime, which is an artifact of the MC simulation's sampling rather than a physical effect. The short baseline relative to the length of the decay volume renders the detector particularly sensitive to the lifetime of the dark photon. When combined with the necessity of simulating the effects of the KMAG resulted in some visible statistical noise in the sensitivity contours.
In Fig. \ref{fig:dp_update_b}, we show the sensitivity projections of the $10^{20}$ POT runs for SQ/DQ and LongQuest-I.
Here, we demonstrate the potential advantage of such an improved decay detector and the reduction of background rate to a 10 percent-level. 
Realistic background estimations of both setups are over the scope of this paper.

%Fluka, Geant-4

\section{Future Direction}\label{sec:future}

In this article, we analyzed the high-energy and high-intensity experiments optimized for the study of long-lived decaying particles, which can be referred to as "decay detectors," including CHARM DD, NuCal (beam-dump run), NA62  (beam-dump run), SQ/DQ, and the LongQuest upgrades (although some of them have other purposes).
The common features of beam dump-type decay detectors are: 
\begin{itemize}
    \item large decay volume 
    \item low density (low background from SM interactions)
    \item simple design thus relatively low-cost \\ (e.g., instrumented stations for tracking and calorimeter).
\end{itemize}
Sometimes, external magnetic fields are used to separate charged particle pairs or shield the decay volume from muons. 

There is also a set of "scattering detectors," often designed to study neutrino scattering and oscillation. They can be used to study particle decay but are not optimized for it. These experiments include MINERvA \cite{Aliaga:2013uqz}, MiniBooNE \cite{Aguilar-Arevalo:2017mqx}, SBND \cite{Machado:2019oxb}, MicroBooNE \cite{Acciarri:2016smi} and DUNE Near Detector (which also could be multi-purpose depending on the final design) \cite{Abi:2018dnh},  which have also been used to study new physics scenarios (see, e.g., \cite{deNiverville:2011it, Kahn:2014sra, Pospelov:2017kep,Magill:2018jla,Magill:2018tbb,Arguelles:2018mtc,Arguelles:2019xgp}). They usually have a smaller volume, higher density, and often a more complicated design compared to the decay detectors (and thus higher costs).
These detectors can potentially provide constraints and novel sensitivity to iDM in different regions of parameter space. We leave these for a future study \cite{New_iDM_paper}.

Future experiments like SHiP \cite{Alekhin:2015byh} and DUNE \cite{Abi:2018dnh} could both potentially provide improved sensitivity to some of the parameter space we explored in this paper. However, since their planned installation is on a much longer timescale than the other experiments we considered, we intend to conduct detailed studies after the designs are finalized.

Finally, we discussed the prospects of the SQ/DQ experiment and the LongQuest setup proposed in Sec. \ref{sec:LongQuest}.
We regard the LongQuest upgrade as one of the ongoing efforts of a complete reconsideration of many fixed-target facilities to study dark sector particles, including the BNB facilities \cite{Aguilar-Arevalo:2017mqx,Machado:2019oxb}, the SeaQuest/SpinQuest facility \cite{SeaQuest_slides,Gardner:2015wea,Berlin:2018pwi}, and the NuMI facility \cite{Kelly:2018brz,FerMINI_Proposal}. A comprehensive plan for these retooled experiments is currently under study, and LongQuest provides one of the most exciting opportunity to study a wide range of dark sector particles with either sizable or small couplings to the SM sector. The exploration of other hidden sector interactions (the Higgs portal, neutrino portal, and the higher-dimensional axion portal \cite{Alexander:2016aln,Battaglieri:2017aum}) as well as other new physics signatures at LongQuest will be discussed in \cite{LongQuest_Proposal}.

\section{Acknowledgement}

In memory of Ann E. Nelson, who has encouraged and continues to inspire us. 

We thank Asher Berlin and Gopolang Mohlabeng for helpful correspondence. 
We thank Paddy Fox, Kun Liu, and Sho Uemura for the discussion of the SeaQuest/DarkQuest experimental configurations.
We thank Peter Cooper, Babette Dobrich, and Gaia Lanfranchi for useful discussions regarding the NA62 experiment.
We thank  Carlos Arguelles, Nikita Blinov, Iftah Galon, Matheus Hostert, and Seodong Shin for useful discussions.
The authors thank APCTP (Korea) and TRIUMF (Canada) for stimulating dark matter workshops. YDT sincerely thank CERN, where this work was initiated, and its theory group for the hospitality.

Parts of the work was performed at the Aspen Center for Physics, which is supported by National Science Foundation grant PHY-1607611. The work of PD was supported by IBS (Project Code IBS-R018-D1).
This manuscript has been authored by Fermi Research Alliance, LLC under Contract No. DE-AC02-07CH11359 with the U.S. Department of Energy, Office of Science, Office of High Energy Physics. The United States Government retains and the publisher, by accepting the article for publication, acknowledges that the United States Government retains a non-exclusive, paid-up, irrevocable, world-wide license to publish or reproduce the published form of this manuscript, or allow others to do so, for United States Government purposes.

\bibliography{biblio}
\newpage
\appendix

\section{$\chi_2 \to \chi_1 \ell^+ \ell^-$ 3-body decay}
\label{app:chi2dec}

\begin{figure}[h]
\centerline{\includegraphics[width=0.35\textwidth]{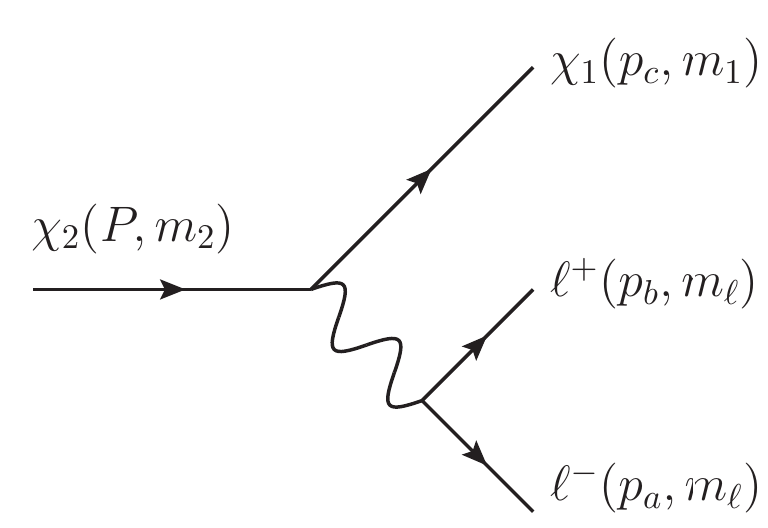}}
\caption{The Feynman diagram for the decay process $\chi_2 \to \chi_1 \ell^+ \ell^-$ with momentum and mass assignments.}
\label{fig:chi2dec}
\end{figure}

Here we show the full differential width of the three-body decay process $\chi_2 \to \chi_1 \ell^+ \ell^-$
\begin{equation}
    \frac{d\Gamma}{dm_{ab}^2dm_{bc}^2} = \frac{1}{(2\pi)^3} \frac{\overline{|\mathcal{M}|^2}}{32 m_2^2},
\end{equation}
where $m_{ij}^2 \equiv (p_i + p_j)^2$ and the amplitude is
\begin{align}
    \frac{1}{4}\Sigma{|\mathcal{M}|^2} & = -\frac{64 \pi^2 \alpha_D \alpha_\mathrm{EM} \epsilon^2}{(m_{ab}^2-m_{A^\prime}^2)^2+\Gamma_{A^\prime}^2 m_{A^\prime}^2} \nonumber \\ 
    \times \bigg[&m_{ab}^4-m_2^2(m_{ab}^2+2m_{bc}^2-2m_1^2) \nonumber \\
    +&2 m_1 m_2 (m_{ab}^2+2m_\ell^2) + m_{ab}^2 (2m_{bc}^2-m_1^2) \nonumber \\
    +&2(m_{bc}^4-m_{bc}^2(m_1^2+2m_\ell^2)+m_\ell^4)\bigg].
\end{align}
The momentum assignments are as shown in Fig. \ref{fig:chi2dec} and $\Gamma_{A^\prime}$ is the width of the dark photon. 

\end{document}